\documentstyle[epsfig]{l-aa}
\begin{document}
\thesaurus{02.04.01, 08.14.1, 08.16.6, 02.07.01}
\title{Rapid uniform rotation of protoneutron stars}

\author{J.O.~Goussard\inst{2} \and P.~Haensel\inst{1,2} 
\and J.L.~Zdunik\inst{1}}

\institute{N. Copernicus Astronomical Center, Polish
           Academy of Sciences, Bartycka 18, PL-00-716 Warszawa, Poland
\and
D{\'e}partement d'Astrophysique Relativiste et de Cosmologie, 
     UPR 176 du CNRS, Observatoire de Paris, 
Section de Meudon, 
 F-92195 Meudon Cedex, France \\
{\em e-mail : goussard@obspm.fr, haensel@camk.edu.pl, jlz@camk.edu.pl} }
\offprints{J.O. Goussard}
\date{}
\maketitle
%\markboth{J.O. Goussard et al.: Rapid rotation of protoneutron stars}
%
%==========================================================
\begin{abstract}
Rapid uniform rotation of newborn neutron stars (protoneutron stars) is 
studied for a range of internal temperatures and entropies 
per baryon predicted by the existing numerical simulations. 
Calculations are performed using general relativistic equations
of hydrostatic  equilibrium of rotating, axially symmetric
stars. 
 Stability  of rotating configurations  with respect to mass
shedding and the axially symmetric perturbations is studied. 
Numerical calculations are performed for a realistic dense matter
equation of state, under various assumptions concerning neutron
star interior (large trapped lepton number, 
no trapped lepton number, 
isentropic,  isothermal).
For configurations with baryon mass well below the maximum
one for the non-rotating models, the mass shedding limit depends
quite sensitively on the position of the 
 ``neutrinosphere'' (which has a deformed, spheroidal shape); 
 this
dependence weakens with increasing baryon mass. 
The absolute upper limit on rotation frequency is, to a good
approximation,  obtained for the maximum baryon mass of
rotating configurations. Empirical formula for the maximum
rotation frequency of uniformly rotating protoneutron stars  
is shown to be quite precise; it actually coincides with
that used for  cold neutron stars.  
Evolutionary sequences at fixed baryon
mass and angular momentum, which correspond to evolution of
protoneutron stars into cold neutron stars are studied, and
resulting constraints on the maximum rotation frequency of
solitary pulsars are discussed. 

\end{abstract}
%

%%%%%%%%%%%%%%%%%%%%%%%%%%%%%%%%%%
\section{Introduction }
Neutrons stars are born in gravitational collapse of massive,
degenerate stellar cores. Newly born neutron stars are hot and
lepton rich objects, quite different from ordinary low
temperature, lepton poor neutron stars. In view of these
differences, newly born neutron stars  are called 
{\it protoneutron} stars; they transform into standard neutron stars on a
timescale of the order of ten seconds, needed for the 
loss of a significant
lepton number excess via emission of neutrinos trapped in the dense, hot
interior. 

In view of the fact that the typical evolution timescale of
a protoneutron star (seconds) is some three orders of magnitude
longer, than the dynamical timescale for this objects (milliseconds), 
one can study  its evolution in the quasistatic approximation
(Burrows \& Lattimer 1986). Properties of static (non-rotating)
protoneutron stars, under various assumptions concerning
composition and equation of state (EOS) of hot, dense stellar interior
 were studied by numerous authors (Burrows \& Lattimer 1986,
Takatsuka 1995, Bombaci et al. 1995, Bombaci 1996, 
Bombaci et al. 1996). 

The scenario of transformation of a protoneutron star  into a
neutron star could be strongly influenced by a phase transition
in the central region of the star. Brown and Bethe (1994)
suggested a phase transition implied by the $K^-$ condensation
at supranuclear densities. Such a $K^-$ condensation could
dramatically soften the equation of state of dense matter,
leading to a low maximum allowable mass of neutron stars. 
In such a case, the massive protoneutron stars could be
stabilized by the effects of high temperature and of the
presence of trapped neutrinos, and this would lead to maximum
baryon mass of protoneutron star larger by some $0.2~M_\odot$ than
that of cold neutron stars. The deleptonization and cooling of
 protoneutron stars of baryon mass exceeding the maximum
allowable baryon mass for neutron stars,   would  then 
inevitably 
lead to their
collapse  into black holes. The dynamics of such a process was
recently studied by Baumgarte et al. (1996). It should be
mentioned, however, that the very possibility of existence of
the kaon condensate  (or other exotic phases of matter, such as
the pion condensate, or the quark matter) at neutron star
densities is far from being  established. Recently, for
instance, 
Pandharipande et al. (1995) pointed out that kaon-nucleon and
nucleon-nucleon correlations in dense matter raise
significantly the threshold density for kaon condensation,
possibly to the densities higher than those characteristic of
stable neutron stars. In view of these uncertainties, we will
restrict in the present paper to a standard model of dense
matter, composed of nucleons and leptons. 

The calculations of the static models of protoneutron stars should be
considered as a first step in the studies of these objects. It
is clear, in view of the dynamical scenario of their formation,
that protoneutron stars are far from being static. Due to the
nonzero initial angular momentum of the collapsing core,
protoneutron stars 
are expected to rotate. On the other hand, the formation scenario
involves  compression (with overshoot of central density) and a 
hydrodynamical bounce, so that a newborn protoneutron star 
begins its life in a highly excited state, pulsating around its
quasistatic equilibrium. 
In the present paper we study  the rotation of
protoneutron stars; pulsations of protoneutron stars will be
discussed in a separate paper (Gondek, Haensel \& Zdunik, in
preparation).

Some aspects of  rapid uniform rotation of protoneutron stars have been
recently studied in (Takatsuka 1995, Hashimoto et al. 1995). 
 However, the calculations reported by Takatsuka (1995) were
actually done for static
(non-rotating) protoneutron stars, and were then used to estimate
the maximum rotation frequency of  uniformly rotating 
protoneutron stars,
$\Omega_{\rm max}$,  via an 
``empirical formula''.  It should be stressed, that the validity
of such an ``empirical formula'', which expresses $\Omega_{\rm
max}$ in terms of the mass and radius of the extremal {\it
static} configuration with maximum allowable mass, had been
checked only in the restricted case of {\it cold} neutron stars
(Haensel \& Zdunik 1989, Friedman et al. 1989, Shapiro et al. 1989, 
Haensel et al. 1995, Nozawa et al. 1996). Only isentropic 
equations of state were 
considered by Takatsuka (1995). 
Hashimoto et al. (1995) calculated the structure of
stationary configurations of uniformly rotating protoneutron
stars, using a two-dimensional general relativistic code.
These authors  restricted themselves to the case with zero
trapped lepton 
number. They assumed a constant temperature 
in the hot interior of the star, 
 and used a zero temperature (cold) 
EOS for  $\rho<10^{10}~{\rm g~cm^{-3}}$.  
It should be stressed, that the
assumption  of $T=const.$  
 corresponds to an isothermal state in the Newtonian (flat
space-time) theory of gravitation. 
In general relativity, we will define isothermal state  by 
 $T^*={N\over \Gamma}T=const.$ 
(where  $N$ is the lapse function and $\Gamma$ is the Lorentz factor, 
see Section 3.1), and the effects of the
space-time curvature will turn out to be rather important for massive
neutron stars. Also, their choice for the low density edge of
the hot interior can be questioned. Finally, 
their criterion for finding maximally rotating configuration is
actually valid only for cold ($T=0$) or isentropic protoneutron
stars: its use in the case of the $T=const.$ hot interior is
unjustified (see Section 3.2 for a correct statement of the
stability criterion). 
   In a recent paper, Lai and Shapiro (1995) 
have studied the secular evolution, secular ``bar instability'', 
and the gravitational wave emission from the newly formed, 
rapidly rotating neutron stars. However, these authors used 
unrealistic (polytropic) equations of state of neutron star 
matter. Moreover, the calculations were done within Newtonian 
theory of gravitation.  In view of this, the internal structure 
 of their models of newly born neutron stars was quite different 
from that characteristic of the realistic models of 
protoneutron stars. The problem of the secular ``bar instability'' 
in rapidly rotating neutron stars was also studied, using general 
relativity, by Bonazzola et al (1995). However, numerical 
calculations were done only for realistic equations of state of 
{\it cold}  neutron star matter.

In the present paper we study the properties 
of uniformly rotating protoneutron stars, using exact 
relativistic description of the
rapid, stationary rotation, combined with  realistic equations
of state of hot dense matter, used in the whole range
of temperatures and densities relevant for protoneutron stars. 
  In particular, we calculate the
maximum  frequency of  
uniform rotation of protoneutron stars and its
dependence on their baryon mass, and  on the thermal state and
composition of stellar interior. It is clear, that uniform rotation 
represents only an approximation to the actual rotational state of 
a newly born protoneutron star. Existing numerical simulations of 
gravitational collapse  of rotating cores of massive stars produce 
differentially rotating protoneutron stars (Janka \& Moenchmeyer 
1989a, b, Moenchmeyer \& Mueller 1989).  However, it should be 
stressed that the initial rotational state of collapsing core is 
unknown, and this implies uncertainty  concerning the rotational 
state of resulting protoneutron star. It is reasonable to say, that 
 the actual degree 
of nonuniformity of rotation  of a protoneutron star 
should be considered as unknown.  In the present paper we will not 
address the question of the physical mechanisms that could ``rigidify'' 
the rotational motion within the protoneutron star interior. However, 
we will use the approximation of uniform rotation in order to limit 
the number  the parameter space for our numerical calculation, and 
also because of the relative simplicity of the stability analysis 
in this specific, idealized case.

 Within our simplified model, the  ``neutrinosphere'' 
 (which has actually a deformed, spheroidal shape) will separate
hot, neutrino-opaque  interior of a protoneutron star 
(hereafter referred to as ``hot interior'') from a
significantly cooler, neutrino-transparent envelope. 
The actual thermal state of the hot interior of protoneutron star is
determined by its formation scenario, and is expected to be
influenced by the dissipative processes (damping
of pulsations, viscous damping of differential rotation, neutrino
diffusion). For simplicity, we will restrict ourselves to two
limiting cases: an isothermal 
($T^*={N\over \Gamma}T=const.$, see
Section 3), and an
isentropic (entropy per baryon 
 $s=const.$) hot  interior. We will also 
consider two limiting cases of the lepton composition of the
protoneutron star interior.  The first case, referring to the
very initial state of protoneutron star,  will correspond to a
fixed trapped lepton number. In the second case, neutrinos will
not contribute to the lepton number of the matter, which will
correspond to vanishing chemical potential of the electron
neutrinos; such a situation will take place after
a deleptonization of a protononeutron star. The position of
the neutrinosphere will be located using  a simple prescription
based on specific properties of  the neutrino opacity of hot
dense matter. In all cases, the equation of state of hot dense 
matter will be determined using one of the models of
Lattimer and Swesty (1991).

The plan of the paper is as follows.  
 In Section 2 we describe
the physical state of the interior of protoneutron star, with
particular emphasis on the EOS of the hot interior at various
stages of evolution. We explain also our
prescription for locating the 
 ``neutrinosphere'' of a protoneutron star, and we  
give  some details concerning the assumed temperature profile within
 a protoneutron star. 
Using simple estimates of the timescales
relevant for various transport processes, we justify the
approximation of stationarity which  is used throughout this
paper. 
In Section 3 we give a brief description of the exact equations, 
used for the calculation of stationary configuration of uniformly rotating
protoneutron stars. We discuss also stability of rotating
configurations with respect to the axially-symmetric perturbations. 
 The numerical method, used for the calculation of rapidly
rotating configurations of protoneutron star, is briefly described in Section
4, where we also discuss  numerical precision of our
solutions.  Maximum rotation frequency, for various physical
conditions prevailing in the hot stellar interior, 
calculated as a function of the
 baryon (rest) mass of  protoneutron star, is presented in Section 5. 
Then, in Section 6  we
show the validity of an empirical formula, which enables one to
express with a surprisingly high precision the maximum frequency
of  rotating protoneutron stars in terms of the mass and radius of
the maximum mass configuration of static 
(non-rotating) protoneutron stars with
same EOS.  In Section 7 we
study the evolutionary transformation of a rotating protoneutron star into a
cold neutron star. We show that, at fixed rest mass and angular
momentum,  maximum rotation
frequency of protoneutron stars imposes severe  constraints on the
rotation frequency of solitary neutron stars. Finally, Section 8 contains
discussion of our results and conclusion.
%
%%%%%%%%%%%%%%%%%%%%%%%%%%%%%%%%%%%%%%%%%%%%%%%%%%%%%%%%%
\section{Physical state of the interior of protoneutron stars}
We consider a protoneutron star (PNS) 
just after its formation. We assume it has a
well defined ``neutrinosphere'', which separates a hot,
neutrino-opaque interior from colder, neutrino-transparent outer
 envelope. Important parameters, which determine the local state
of the matter in the hot interior are: baryon
(nucleon) number 
density $n$,  net electron fraction $Y_e
= (n_{e^-}-n_{e^+})/n$, and the net electron-neutrino
fraction $Y_{\nu}=Y_{\nu_e}-Y_{\bar\nu_e}$. The calculation of
the composition of hot matter and of its EOS is  described below.
%%%%%%%%%%%%%%%%%%%%%%%%
\subsection{Neutrino opaque core with trapped lepton number}
Such a situation is characteristic of
 the very initial stage of existence
of a PNS. Matter is composed of nucleons (both free and bound in
nuclei) and leptons (electrons and neutrinos; for simplicity, we
do not include muons). All constituents
of the matter (plus photons) are in thermodynamic equilibrium at
given values of $n$, $T$ and $Y_l=Y_e+Y_\nu$. The
composition of the matter is calculated from the condition of
 beta equilibrium, combined with the condition of a fixed
$Y_l$, 
%%%%%%%%%
\begin{eqnarray}
\mu_p + \mu_e &=& \mu_n + \mu_{\nu_e}~,\nonumber \\
Y_l&=&Y_e+Y_\nu~,
\label{mu.trapL}
\end{eqnarray}
%%%%%%%%%%%%%%%
where $\mu_{\rm j}$ are the chemical potentials of matter
constituents. At the very initial stage we expect $Y_l\simeq
0.4$. Electron neutrinos are degenerate, with $\mu_{\nu_e}\gg T$
(in what follows we measure $T$ in energy units). The
deleptonization, implying the decrease of $Y_l$, 
 occurs due to diffusion of neutrinos outward
(driven by the $\mu_{\nu_e}$ gradient), on a timescale of 
seconds (Sawyer \& Soni 1979, Bombaci et al. 1996). 
The diffusion of highly degenerate neutrinos from the central
core is a dissipative
process, resulting in a significant {\it heating} of the
neutrino-opaque core (Burrows \& Lattimer 1986). 
%%%%%%%%%%%%%%%%%%%%%%%%
\subsection{Neutrino opaque core with $Y_\nu = 0$}
This is the limiting case, reached after complete deleptonization. 
There is no trapped lepton number, so that $Y_l=Y_e$ and
$Y_{\nu_e}=Y_{\bar\nu_e}$, and therefore 
$\mu_{\nu_e}=\mu_{\bar\nu_e}=0$. Neutrinos trapped within the
hot interior do not influence the beta equilibrium of nucleons,
electrons and positrons, and  for given $n$ and $T$ the
equilibrium value of $Y_e$ is determined from
%%%%%%%%%
\begin{eqnarray}
\mu_p + \mu_e &=& \mu_n~,
\label{mu.free}
\end{eqnarray}
%%%%%%%%%%%%%%%
while $\mu_{e^+}=-\mu_e$. In practice, this approximation can
be used  as soon as electron neutrinos become non-degenerate
within the opaque core, 
 $\mu_{\nu_e}< T$, which occurs after  some $\sim 10$ seconds
(Sawyer \& Soni 1979, Bombaci et al. 1996). The neutrino
diffusion is then driven by the temperature gradient, and the
corresponding timescale of the heat transport (PNS cooling) can
be estimated as $\sim 50 (T/10~{\rm MeV})^{-3}$~s (Sawyer \& Soni
1979). 
%%%%%%%%%%%%%%%%%%%%%%%%%
\subsection{Neutrinosphere and temperature profile}
In principle, the temperature (or entropy per nucleon) profile
within a PNS has to be determined via evolutionary calculation,
starting from some initial state, and taking into account
relevant transport processes in the PNS interior, as well as
neutrino emission from PNS.  Transport
processes within neutrino-opaque interior occur on timescales of
seconds, some three orders of magnitude longer than dynamical
timescales. The very outer layer of PNS becomes rapidly
transparent to neutrinos, deleptonizes, and cools on a very
short timescale 
via  $e^-e^+$ pair annihilation and plasmon decay. 
It seemed thus natural to model the thermal structure of the PNS
interior by a hot core limited by a ``neutrinosphere'', and an outer
envelope of~ $T<0.5$ MeV. The transition through the ``neutrinosphere''
is accompanied by a temperature drop, which takes place over
some interval of density just above the ``edge''of the hot
neutrino-opaque core, situated at some $n_\nu$. 

In view of the uncertainties in the actual temperature profiles
within the hot interior of PNS, we considered two extremal
situations for $n>n_\nu$, 
corresponding to an isentropic and an isothermal hot interior. In the
first case, hot interior was characterized by a  constant entropy
per baryon $s=const.$. In the case of trapped lepton number, 
this leads to the EOS of the type: pressure 
$p=p(n,~[s,Y_l])$, energy density  
$e=e(n,~[s,Y_l])$, and temperature 
$T=T(n,~[s,Y_l])$, 
 with  fixed  $s$ and $Y_l$. 
 This EOS  will be denoted by EOS[$s,Y_l$]. In the case of
an isentropic, zero trapped lepton number EOS,  we will 
have EOS[$s, Y_\nu=0$]. 

The condition of isothermality, which
 corresponds to thermal equilibrium,  is more
complicated. Due to the curvature of the space-time within PNS, 
the condition of isothermality corresponds to 
the constancy of $T^*={N\over \Gamma}T$ 
(see Section 3.1). The significance of the $T^*=const.$ 
condition will be discussed in Section 3.1. 
In the static case, 
 the isothermal state
within  the hot interior will be reached on a timescale
 corresponding to thermal equilibration, which is much longer
than the lifetime of a PNS. In the case of a rotating PNS, the
situation can be expected to be even more complicated (see,
e.g.,  Chapter 8 of Tassoul (1978)). 
Nevertheless, we considered the
$T^*=const.$ models for several reasons. First, as a limiting case
so different from the $s=const.$ one, it enables us to check the
dependence of our results for rapidly rotating PNS on the
thermal state of the hot interior. 
 Moreover, for the isothermal PNS  we can apply
the  criterion of
stability  with respect to the axi-symmetric perturbations,
and this  will enable us to calculate the value of $\Omega_{\rm
max}$ for the stable supramassive rotating PNS models with an 
isothermal interior (see Section 3).

Our calculation of the ``neutrinosphere'' within the hot PNS
interior was done using a simple method, described below. 
 For a given {\it static} PNS model, the neutrinosphere radius, 
$R_\nu$, has been
located through the condition
%%%%%%%%%
\begin{equation}
\int_{R_\nu}^R
{1\over \lambda_\nu(E_\nu)}
\sqrt{\vert g_{rr}\vert}{\rm d}r= 1~,
\label{R_nu}
\end{equation}
%%%%%%%%%%%%%%%
where $\lambda_\nu$ is the neutrino mean free path which  
is calculated at the matter temperature $T$, 
 while  $E_\nu$ is the mean energy of non-degenerate neutrinos at 
  (and above) the 
neutrinosphere, $E_\nu=3.15T_\nu$. 
  We assumed that opacity above $R_\nu$
is dominated by the elastic scattering off nuclei and nucleons,
so that $\lambda_\nu = \lambda^0_\nu(n,T)/E_\nu^2$. Then,
we determined the value of the density at the neutrinosphere, 
 $n_\nu$,  for a given static PNS model, combining  Eq. (\ref{R_nu})
with that of hydrostatic equilibrium, and readjusting
 accordingly the temperature profile  within the outer layers of
PNS. This value of $n_\nu$ was then used in the calculations of
rotating PNS  models.  
Let us notice, that similar approximation, in which the 
``neutrinosphere'' in rapidly rotating protoneutron star was defined as 
 a surface of constant density, was used  by Janka and Moenchmeyer 
 (1989a, b). Some details concerning actual calculation of the
temperature profile in the vicinity of the ``neutrinosphere'' will
be given in subsection  3.3. 

%%%%%%%%%%%%%%%%%%%%%
\subsection{Equation of state and static
models of PNS}

The starting point for the construction of our EOS for the PNS
models was the model of hot dense matter of Lattimer and Swesty
(1991), hereafter referred to as LS. Actually, we used one
specific LS model, corresponding to the incompressibility
modulus at the saturation density of symmetric nuclear matter
$K=220~$MeV. For $n>n_\nu$ we supplemented the LS model
with contributions 
resulting from the presence of trapped neutrinos of three
flavours (electronic, muonic and tauonic) and of the 
corresponding antineutrinos. 

%
%%%%%%%%%%%%%%%%%
\begin{figure}     %1a
\epsfig{figure=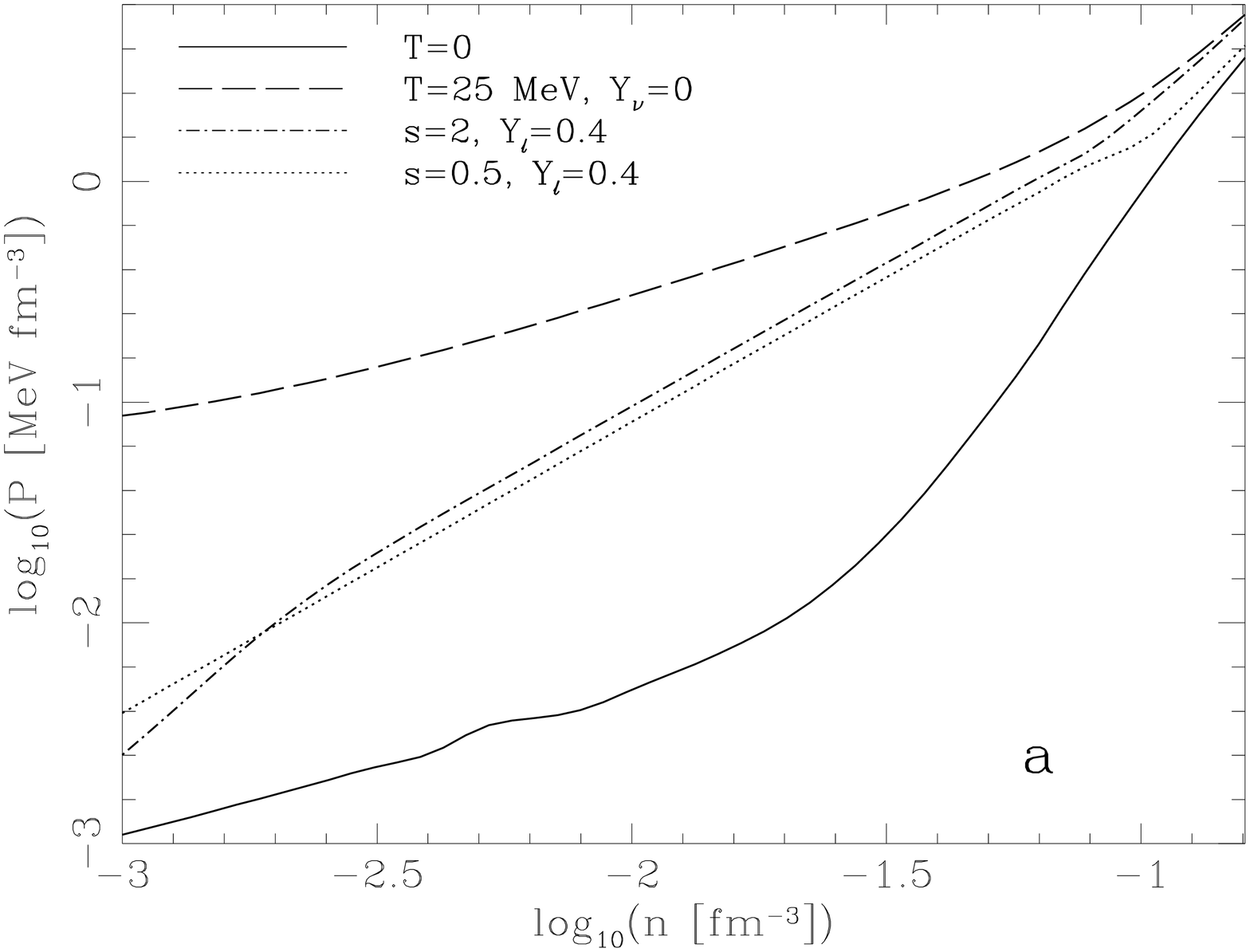,height=8.5cm,angle=-90}
\caption[]{{\bf a.}
Pressure versus baryon density for our model of dense hot matter, 
under various physical conditions, 
for the subnuclear densities ($n<0.16~{\rm fm^{-3}}$).
The curve $T=0$ corresponds to cold catalyzed matter.  
The curve corresponding to $s=0.5, Y_l=0.4$ is unphysical, but 
has been added in order to  visualize the importance of trapped lepton 
number at subnuclear densities.  
}
\addtocounter{figure}{-1}
\end{figure}
%%%%%%%%%%%%%%%%%%%%%%%%
%
%%%%%%%%%%%%%%%%%
\begin{figure}     %1b
\epsfig{figure=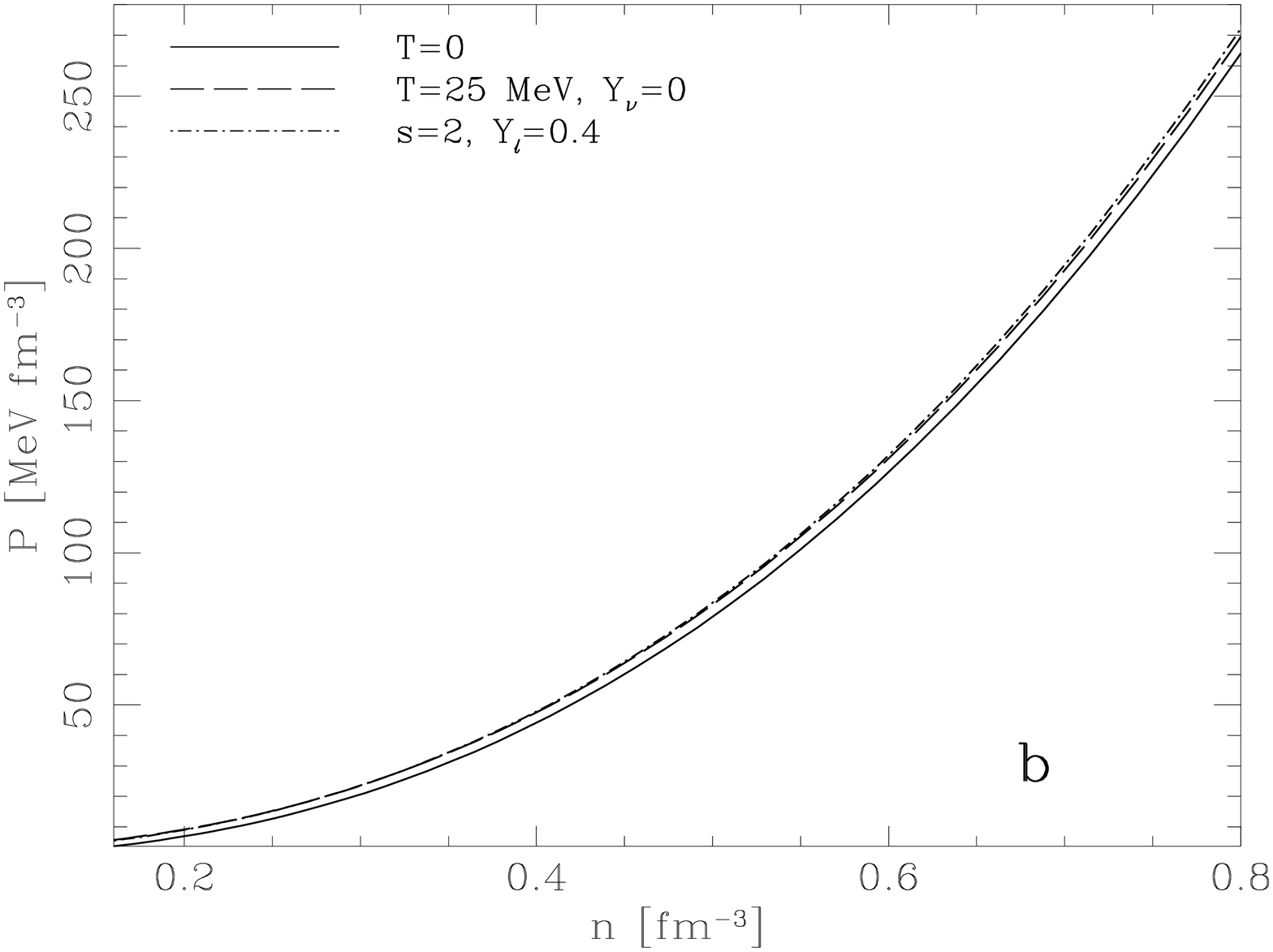,height=8.5cm,angle=-90}
\caption[]{{\bf b.}
Pressure versus baryon density for our model of dense hot matter, 
under various physical conditions, for supranuclear 
densities ($n>0.16~{\rm fm^{-3}}$). Notation as in Fig. 1a. 
}
\end{figure}
%%%%%%%%%%%%%%%%%%%%%%%%

In Fig. 1a, b we show our EOS in several cases, corresponding to
various physical conditions in the hot, neutrino-opaque interior
of PNS. For the sake of comparison, we have shown also the EOS
for cold catalyzed matter, used for the calculation of the
(cold) NS models. In Fig. 1a we show EOS at subnuclear
densities. At these densities, both the temperature and the
presence of trapped  neutrinos stiffen the EOS, as compared to
the cold catalyzed matter one, and the stiffening is rather
dramatic.  The constant $T$ EOS stiffens considerably at lower
densities, which is due to the  weak dependence of
the thermal contribution (photons, neutrinos) on the baryon
density of the matter (this effect will be to some extent
moderated by the factor ${\Gamma\over N}$ in the isothermal
PNS, see Section 3.1).  It is quite obvious, that $T=const.$ EOS 
becomes dominated by thermal effects below for $n<10^{-2}~{\rm
fm^{-3}}$.  On the contrary, for isentropic EOS, the effect of
the trapped 
lepton number ($Y_l=0.4$) turns out to be much more important
than the thermal effects. This can be see in Fig. 1a, 
by comparing dash-dotted curve, 
$[s=2,~Y_l=0.4]$, with the dotted line, which
corresponds to an artificial (unphysical) case with small
thermal effects, $[s=0.5,~Y_l=0.4]$.

It is clear, that the correct location of the 
``neutrinosphere'', which separates hot interior from the colder
outer envelope, should be important for the determination of the
radius of PNS, and in consequence, of the maximum rotation
frequency of a given PNS model. 

It may be useful to compare our subnuclear EOS for PNS with those used by
other authors. 
 The subnuclear EOS of PNS, used in
the papers of Hashimoto et al. (1995) and Bombaci et al.
(1995, 96) is very different from that used in the present 
paper. 
 In particular, 
Hashimoto et al. (1995) 
 used a cold ($T=0$)
EOS for  the densities below $10^{10}~{\rm g~cm^{-3}}$. 
  On the other hand, Bombaci et al. (1995, 96) stop their hot EOS
at the edge of the liquid interior, and use the $T=0$ (cold
catalyzed matter) EOS for the densities below $0.08~{\rm
fm^{-3}}$; in this way, they seriously underestimate 
thermal and neutrino trapping effects on the radius of PNS.  

Our EOS above nuclear density are plotted in Fig. 1b. The
presence of a trapped lepton number softens the EOS, while
thermal effects always stiffen it with respect to that for cold
catalyzed matter. The softening of the supranuclear EOS at fixed
$Y_l$ is due to 
the fact, that a significant trapped lepton number increases the
proton fraction, which implies the softening of the nucleon
contribution to the EOS.

It should be stressed, that in contrast to Hashimoto et al.
(1995) and Bombaci et al. (1995, 96) we used  a 
unified dense matter model, valid for both 
supranuclear and subnuclear densities. Also, the fact that we
use various assumptions about the $T$ and $s$ profiles within
PNS, enables us to study the relative importance of the 
 temperature profile
and that of a trapped lepton number, for the PNS models.

%%%%%%%%%%%%%%%%%
\begin{figure}     %2
\epsfig{figure=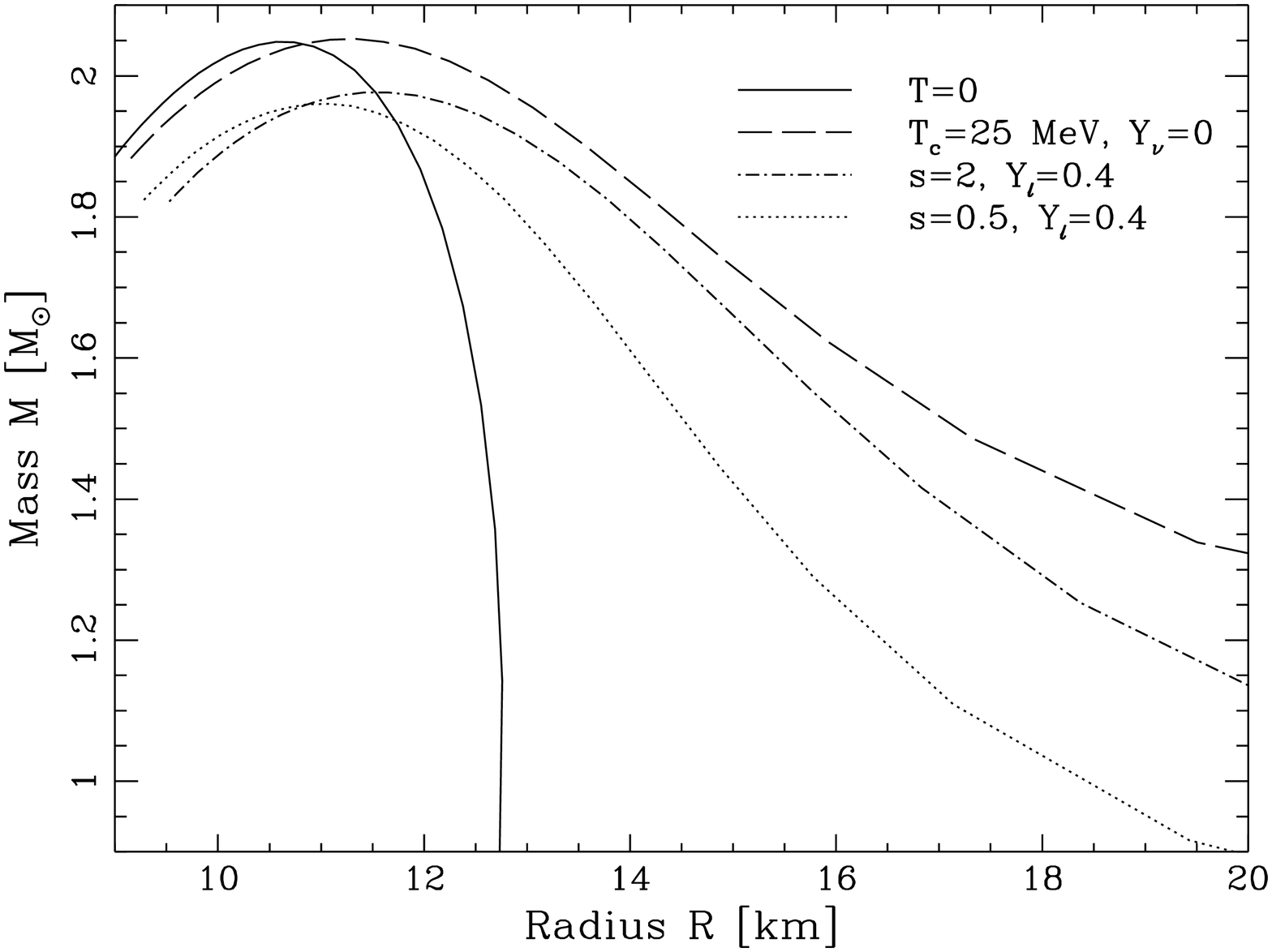,height=8.5cm,angle=-90}
\caption[]{
The gravitational mass versus stellar areal radius for static models 
of the protoneutron stars and neutron stars, under various assumptions 
concerning the physical conditions within the stellar interior. 
The curve corresponding to $s=0.5, Y_l=0.4$ is unphysical, but has been 
added in order to visualize the relative   importance of the trapped lepton 
number and thermal effects. The curve $T=0$ corresponds
to cold neutron stars. The curve $T_c = 25~ {\rm MeV}$ includes the
${\Gamma\over N}$ factor.
}
\end{figure}
%%%%%%%%%%%%%%%%%%%%%%%%

The mass-radius relation for the static PNS models calculated using
various versions of our EOS for the hot interior is shown in
Fig. 2.  We assumed $n_\nu = 5\times 10^{-3}~{\rm fm^{-3}}$,
which was consistent with our definition of the ``neutrinosphere''. 
For the sake 
of comparison, we show also the mass-radius relation for the
$T=0$ (cold catalyzed matter) EOS, which corresponds to cold 
neutron star models. In the case of the isothermal hot interior
with central temperature 
$T_{\rm c}=25$~MeV  we note a very small increase of
the maximum mass, as compared to the $T=0$ case (c.f., Bombaci
et al. 1995, 1996). However, the
effect on the mass-radius relation is quite strong, and
increases rapidly with decreasing stellar mass. In the case of
the isentropic EOS with a trapped lepton number, [$s=2,Y_l=0.4$],
the softening of the high-density EOS due to the trapped $Y_l$
leads 
to the decrease of $M_{\rm max}$ compared to the $T=0$ case; as far
as the value of $M_{\rm max}$ is concerned, the softening
effect of $Y_l$ prevails over that of finite $s$ (this is
consistent with results of Takatsuka 1995 and Bombaci et al.
1995, 1996). However, the thermal effect on the radius is very 
important even in the case of  $Y_l=0.4$. 
This can be seen by comparing the $[s=2,~Y_l=0.4]$ curve with that
corresponding to the unphysical, fictitious case of
$[s=0.5,~Y_l=0.4]$. 

The very initial state of a PNS corresponds  to a significant
trapped lepton number. With our assumption of a ``standard''
 composition of dense matter (i,e., excluding large amplitude
$K^-$-condensate, or a very large percentage of hyperons in cold
dense matter), the maximum baryon mass (baryon number) 
of PNS is lower than that of
cold NS. Therefore, in our  case a stable PNS transforms into a
stable NS, and  the scenario PNS$\longrightarrow$Black Hole, 
 considered by Baumgarte et al. (1996) is excluded.

At a  given mass, the radius of a PNS is significantly larger
than that of a cold NS. As remarked by Hashimoto et al. (1995),
this should have important implications for rotating 
PNS. It should be stressed, however, that the value of radius,
especially for PNS  which are  not close to the $M_{\rm max}$
configuration, turns out to be 
 quite sensitive to the location of the edge of
the hot neutrino-opaque interior (i.e., to the value of
$n_\nu$). The choice of Bombaci et al. (1996) would lead to a
much smaller effect on $R$, while that of Hashimoto et al.
(1995) would result in larger  values of the
PNS radii. 
%
%%%%%%%%%%%%%%%%%%%%%%%%%%%%
\subsection{Stationarity}
The EOS of PNS is evolving with time, due mainly to the
deleptonization process, which  changes the composition of 
matter, and also due to changes of the internal temperature of the
star. However, these changes occur on the timescales $\tau_{\rm
evol}\sim $1-10~s, which are
three or more orders of magnitude longer than the dynamical
timescale, governing the readjustment of pressure and gravity
forces. This dynamical timescale  $\tau_{\rm dyn}\sim 1~$ms
corresponds also  to the  characteristic periods of the PNS
pulsations and of their rapid rotation. In view of this, we are
able to decouple PNS evolution from its dynamics, and treat its
rotation in the stationary approximation, with a well defined
EOS of the PNS matter. 

One of the neglected dynamical processes, implied by the 
 radiative processes and the evolution of the thermal 
structure of a rotating PNS, is  the meridional circulation of the 
matter. Strictly uniform rotation is incompatible with an 
assumption of a steady thermal state, resulting from  the diffusive 
(radiative) equilibrium (see, e.g., Tassoul 1978). 
The requirement  of radiative equilibrium will necessarily imply 
the existence of a meridional circulation of the matter. 
However, the velocity of this meridional circulation will be 
of the order of the stellar radius divided by the thermal timescale 
(the timescale of changes of the entropy of the PNS interior, 
which is of the order of neutrino diffusion timescale), 
 which is much smaller than the rotational velocity 
(see, eg., Section 8 of Tassoul 1978). 
 In view of this, we can neglect the effect of the meridional 
circulation when calculating the mechanical equilibrium of a rapidly 
rotating PNS. Finally, let us notice that in a special, idealized 
case of $T^*=const$, the radiative flux vanishes. Then, pure 
uniform rotation can be realized as a steady state of a PNS.  

%%%%%%%%%%%%%%%%%%%%%%%%%%%%%%%%%%%%%%%%%%%%%%%%%%%%%%%%%
\section{Formulation of the problem}
The problem of the calculation of the stationary state of
uniform rotation of cold neutron stars, within the framework of
general relativity, was considered by numerous authors (see the
review article of Friedman and Ipser 1992, and references
therein). Extensive calculations for a broad set of realistic
equations of state of cold dense matter were recently presented
in (Cook et al. 1994) and (Salgado et al. 1994). Here, we will
extend the methods used at $T=0$ to the case  of hot PNS. We
will use the notation and formalism developed in the paper of
Bonazzola et al. (1993), hereafter referred to as BGSM. 
\subsection{Equation of stationary motion}
One of the problems introduced by finite temperature is that,
if one does not make  any other assumption about the equilibrium,
the equation of stationary motion does not have a first
integral (Bardeen 1972).

In the notation of BGSM, the equation  of stationary motion  
(Eq. 3.25 of BGSM) reads, when the effects of temperature and rigid 
rotation are included, as
%%%%%%%%%%%%%%
\begin{equation}
{\partial}_i (H + \ln \frac{N}{\Gamma}) = T e^{-H} {\partial}_i s \label{eqsta}
\end{equation}
%%%%%%%%%%%%%%%
where $s$ is the entropy per baryon, $T$ is the 
temperature, and $H = \ln[(e+p)/(n m_0 c^2)]$ is the so-called
pseudo-enthalpy (or log-enthalpy), 
$N$ is the lapse function appearing in the space-time metric, 
and $\Gamma$ is the Lorentz factor due to rotation. 
The equation of the stationary motion is to be supplemented with 
the equations determining the metric functions (see BSGM for the
derivation and the explicit form of the complete set of
equations).

In the expression
for $H$, $e$ is the energy density (which includes rest energy
of matter constituents) and $m_0$ is the nucleon rest mass. 

This equation is the general relativistic equivalent of
a well-known Newtonian formula (see e.g. Tassoul 1978).
It is
straightforward to show that a {\it sufficient} condition for (\ref{eqsta})
 to be integrable is $T = T(n)$.
In this case, one obtains a first integral 
of motion of the form :
%%%%%%%%%%%%%%%%%%%
\begin{equation}
H + \ln \frac{N}{\Gamma} - \int T e^{-H}ds = const. \label{firstint}
\end{equation}
%%%%%%%%%%%%%%%%%%
This enables us to calculate the density profile in the envelope
of the PNS, because we have assumed  a specific  
$T(n)$ profile 
 within  it (see section 3.3).
Let us stress that, due to the integral term in (\ref{firstint}), the
pseudo-enthalphy  is no more an explicit function of the
metric potentials, as it was in BGSM. 
In fact, one must solve (\ref{firstint}) to have $H$ 
within the star.

In the two particular cases,  chosen by us for the hot interior,
 specific first integrals of Eq. (\ref{eqsta}) can be found.  
First, in the case of general relativistic thermal equilibrium,
$T^{\star} = T \frac{N}{\Gamma} = const.$, it is easy to show that
%%%%%%%%%%%%%%%
\begin{equation}
\mu^{\star} = \mu \frac{N}{\Gamma} = const., 
\end{equation}
%%%%%%%%%%%%%%%
where 
$\mu = (e+p)/n - Ts$ is the baryon chemical potential, is
indeed the integral
of (\ref{eqsta}). The constancy of $T^*$ and $\mu^*$ 
corresponds to the general-relativistic thermodynamic
equilibrium, if we neglect the 
time dependence of the EOS of PNS (i.e., if we freeze transport
phenomena). 
 Second, in the case of isentropic profile $s(n,T(n)) = const.$, 
the thermal term in (\ref{eqsta}) vanishes and the first integral 
of Eq. (\ref{eqsta}) reads
%%%%%%%%%%%%%%%
\begin{equation}
T^{\star}s + \mu^{\star} = const.
\end{equation}
%%%%%%%%%%%%%%%
%
%%%%%%%%%%%%%%%%%%%%%%%
\subsection{Instabilities and maximum mass}
The instabilities that could develop in rapidly rotating PNS,
and which could limit  the maximum angular frequency of these
objects, are of secular and of dynamic type. 

Let us start with the problem of secular stability with respect
to  the axi-symmetric perturbation of rotating configurations.  
We denote the total baryon number, total angular momentum, and total
entropy  of a PNS by $N_{\rm bar}$, $J$, and $S$, respectively. 
We used the {\it secular} instability criterion of Friedman
et al. (1988), extended to the case of finite temperature.  
For the purpose of completeness, we restate here their lemma 
(and correct a misprint of their paper).\\

{\bf Lemma :} Consider a three-parameter family of uniformly
rotating hot stellar models having an equation of state of
the form $p=p(e,T)$. Suppose that along a continous sequence 
of models labelled by a parameter $\lambda$, there is a point
 $\lambda_0$ at which $\dot{N}_{\rm bar} = 
 {\rm d}N_{\rm bar}/{\rm d}\lambda$, $\dot{J}$
and $\dot{S}$ vanish and where 
$\frac{d}{d\lambda}(\dot{\mu^{\star}}\dot{N}_{\rm bar} + 
\dot{\Omega}\dot{J}+ \dot{T^{\star}}\dot{S}) \neq 0$. 
Then the part of the
sequence for which 
$\dot{\mu^{\star}}\dot{N}_{\rm bar} + \dot{\Omega}\dot{J}+ 
\dot{T^{\star}}\dot{S} > 0$ is unstable for $\lambda$ near $\lambda_0$.\\

Similarly as in the case of Friedman et al. (1988), this lemma follows
directly from  
Theorem I of Sorkin (1982), with his function $S$ 
replaced by $-M$,
the $E^\alpha$  quantities replaced 
by $N_{\rm bar}$, $J$, $S$ and the $\beta^\alpha$  ones by 
 $\mu^{\star}$, $\Omega$, $T^{\star}$. The conditions of the theorem are
fulfilled because stellar models are configurations for which $M$
is minimized at fixed $N_{\rm bar}$, $J$ and $S$, 
{\it and} because 
difference in $M$ between two neighbouring equilibria  can be
expressed as (Bardeen 1970)
%%%%%%%%%%%%%%%
\begin{equation}
 {\rm d}M =  \mu^{\star} {\rm d}N_{\rm bar}
  + \Omega {\rm d}J + T^{\star}
{\rm d}S .
\end{equation}
%%%%%%%%%%%%%%%%

Let us stress here that this last equation requires either $T^{\star}$ 
or $s$ to be constant through the whole star, which fixes, 
in each of these two cases, 
the temperature profile within the stellar model.  

We investigated the stability of our models with a specific
 version of this criterion, in which we choose a 
continous sequence of equilibria to be a sequence at fixed 
$N_{\rm bar}$ 
and $S$. The point of the loss of stability is then simply
the point of extremal $J$, i.e. :
%%%%%%%%%%%%%%%%%%
\begin{equation}
\left(
\frac{\partial J}{\partial \rho_{\rm c}}
\right)_{N_{\rm bar},S} = 0,
\end{equation}
%%%%%%%%%%%%%%%%%%
where $\rho_{\rm c}$ is the central density of the star.

The instability discussed above is a secular one; it will
develop on the timescale needed to transport the angular
momentum within the perturbed model, in order to decrease
the energy of the star while changing its shape and structure.
However,  the timescale of the PNS evolution is quite short
(seconds), and it is driven by the same transport processes
involving neutrinos, as those which are needed to destabilize
the star via the axi-symmetric perturbations. In view of this,
one might expect that the secular instability described above is
not efficient in disrupting the quasi-stationary configuration
of rapidly rotating PNS. 

However, we should remember that the above considerations apply
to the case of infinitesimal perturbations. Transport processes
would be crucial for removing the energy barrier separating the
initial, secularly unstable configuration, from the dynamically
unstable one, which would eventually collapse into a black hole.
  However, newly born PNS are expected  to be in a highly excited
state, in which various modes of stellar pulsations are
excited. One may expect that the energy contained in these
pulsations is sufficient to overcome the energy barrier
separating the actual metastable, secularly unstable state from
the dynamically  unstable, collapsing one. 
In view of this, we expect  that the
secularly unstable configurations should be treated like unstable
ones. Therefore, the critical configuration,  given by Eq.(9),
will be thus considered as the last stable one.

A rapidly rotating PNS can be also susceptible to other types 
of secular instabilities. The instability with respect to the 
non axisymmetric perturbations can be driven by the
gravitational radiation reaction (GRR). However, detailed
calculations performed for hot NS suggest, that at the
temperature exceeding $10^{10}~$K these instabilities can only
slightly decrease the maximum rotation frequency of uniform
rotation, due to the damping effect of the matter viscosity 
 (Cutler et al. 1990, Ipser \& Lindblom 1991, Lindblom 1995,
Yoshida \& Eriguchi 1995, Zdunik 1996). Secular instabilities
of rapidly rotating NS with respect to the non-axisymmetric
``bar''  mode were recently 
investigated by Lai \& Shapiro (1995) and Bonazzola et al.
(1995). While Lai \& Shapiro (1995) addressed the problem of
``bar'' instability of newly formed NS, they assumed a rather
unrealistic ``ellipsoidal model'', 
involving only shear viscosity as a source of viscous dissipation, 
  and performed their calculations within the Newtonian
theory of gravity. On the other hand, in their relativistic
calculations Bonazzola et al. (1995) considered
only cold NS, and found that the secular ``bar'' instability can
set in before the Keplerian (mass shedding) limit is reached
only for sufficiently stiff EOS of NS matter. 
 The problem of the ``bar'' instability of PNS, 
 with realistic EOS of the hot interior, will be investigated 
by us in the future. 
 In any case, 
inclusion of possible additional secular instabilities can only
decrease the maximum rotation frequency of PNS below the values
obtained using the simplified approach adapted in the present
paper.

The fact of the existence  of the maximum mass of rotating
PNS,  $M_{\rm max}^{({\rm rot})}$ (and maximum baryon mass, $M_{\rm
bar,max}^{\rm(rot)}$) (see Section 5), 
puts a well defined and stringent limit on
$\Omega_K$ which can be reached by PNS. 
 Configurations with $M_{\rm bar}>M_{\rm bar,max}^{\rm(rot)}$ are
{\it dynamically unstable}: no stationary solution exists above
$M_{\rm bar,max}^{\rm(rot)}$. At the same time, the angular frequency of
rigid rotation cannot exceed the Keplerian value, $\Omega_K$.
These two conditions, combined with the instability criterion,
expressed in Eq. (9), determine the maximum frequency (or,
strictly speaking, an upper bound on the frequency ) of rigid
rotation of PNS. Practical implementation of these criteria will
be  described in Section 5. 
%
%%%%%%%%%%%%%%%%%%%%%%%%%%%
\subsection{Isothermal and isentropic temperature profiles}
A consistent study of PNS would require
an exact treatment of the thermal transport in the frame
of a non-stationary spacetime. Unfortunately, 
such a study is beyond the scope of the present work, 
 and we have thus chosen to impose ``by hand'' the temperature profile
 within the star, following prescription described in  the
subsection  2.3 . We divided the interior of PNS into the
hot interior ($n > n_\nu$), a layer corresponding to the
temperature drop within the ``neutrinosphere'' 
($n_\nu-\Delta n_\nu < n <n_\nu$), and the low
temperature, neutrino-transparent outer envelope with 
$n<n_\nu -\Delta n_\nu$. We have chosen two types of temperature
profiles within the hot, neutrino-opaque core 
 $(n > n_\nu)$ :
\begin{itemize}
\item the isothermal  profile : $T=T^{\star}\frac{\Gamma}{N}$,
 with $T^\star= const.$,
\vskip 3mm
\item the isentropic profile :  $T=T(n)$, with $s(n,T(n))=const$.
\end{itemize}

For the transition region and the low temperature envelope, 
we used a suitable profile $T(n)$ :
\begin{eqnarray}
T(n) &=& 0.2 \; {\rm MeV} \;\: {\rm for} \;\:  
n < n_\nu - \Delta n_\nu \nonumber \\
T(n) &=& f(n) \;\: {\rm for} \;\: 
n_\nu > n > n_\nu - \Delta n_\nu. \nonumber 
\end{eqnarray}
The function  $f$ was chosen  for computational convenience as
a suitable combination of an exponential and a gaussian
function, selected to lead to  $T(n)$ of class ${\cal C}^1$ 
through the transition (temperature drop) region. Our typical
choice was $\Delta n_\nu/n_\nu = 2~10^{-2}$; 
increasing this value up $\Delta n_\nu/n_\nu =  0.2$ led to a
very small increase of the stellar radius.

As the mass of a massive PNS is almost entirely contained
in the hot neutrino-opaque core (a rough estimate for a typical
star gives less than $10^{-3}$ of the total mass for the cool
envelope mass), we supposed that our stability
criterion remained 
valid also in the case of the presence of the low temperature
envelope. 

%
%%%%%%%%%%%%%%%%%%%%%%%%%%%%%%%%%%%%%%%%%%%%%%%%%%%%%%%%%
\section{Numerical method}
We used a code based on the $3+1$ formulation of the Einstein
equations in stationary axisymmetric spacetimes (BGSM). 
The four elliptic
equations obtained were solved by means of a spectral method, in which
the functions are expanded in  different polynomial bases 
(Chebyshev for $r$, Legendre for $\theta$ and Fourier for $\phi$). 
We refer the reader to BGSM for a detailed 
description of the code, including
the description of the ``virial'' indicator used for monitoring
the convergence and the precision reached. Let us just say 
that the code was modified to take into account thermal effects
and, contrary to Salgado et al. (1994), to converge to the 
solution with various 
quantities being held fixed (for example $M_{\rm bar}$ or $J$ or $S$).
Here we briefly outline some of the numerical checks we made.

The two-parameters EOS was interpolated using bicubic splines
 subroutine from the NAG library. In this way, the thermodynamic
functions are of class ${\cal{C}}^{2}$, but the thermodynamic
consistency is not conserved.

The global relative error, evaluated by the means of the virial
check (see BGSM), is $\sim 2 \cdot 10^{-3}$. This relatively ``low''
precision is due to the thermodynamical inconsistencies.

We used two grids in $r$ for the star and one in $1/r$ for the
exterior with $N_r = 33$ in each, and one
grid in $\theta$ with $N_\theta = 17$. We checked that a
greater $N_r$ or $N_\theta$ does not change the results by more
than a few $10^{-4}$ at most, which stays within 
 the global precision reached. Let us stress that such a low number
of grid points is sufficient to reach high accuracy within a
spectral method (that would not be the case with a finite difference
scheme). 

The temperature drop  at the ``neutrinosphere'' is not always located on
the border of the internal grid in $r$, which could influence
the precision (remember that the 
 spectral methods are very sensitive to 
discontinuities). We checked that, in fact, even when the 
 ``neutrinosphere'' is far from the border of the grid, the global
physical quantities of the star did not change much  
(we found also relative variations of a few 
$10^{-4}$ at most).

For a ``simple'' model where $(T_c, \rho_c, \Omega_c)$ 
were held fixed, convergence required  no more than $50$ iterations
and $\sim 1$ minute of CPU time on a Silicon Graphix Indigo$^2$ 
workstation.
 However, the number of iterations can reach $400$ if one needs
to converge    
to, e.g., a Keplerian configuration at fixed supra-massive
baryon mass.

Finally, we compared our results at $T=0$ and $\Omega=0$ with
the results of another code used by one of us (P. Haensel), and
found that the relative differences in the  global properties of
the stellar models are of the order of 
$\sim 10^{-3}$, which can be imputed to the different interpolation 
procedures. 
%
%%%%%%%%%%%%%%%%%%%%%%%%%%%%%%%%%%%%%%%%%%%%%%%%%%%%%%%%%
\section{Numerical results. Maximally rotating protoneutron stars}
For a given EOS, rapidly rotating PNS models can be divided into
two families: normal and supramassive one. Normal models are
those with baryon mass which does not exceed the maximum
allowable baryon mass of static (non-rotating) configurations,
$M_{\rm bar,max}^{({\rm stat})}$.  
A {\it normal} rotating model can be transformed  into a static
configuration of the same baryon mass, through a continuous
decrease of $\Omega$. 
Rotation of PNS increases their maximum allowable mass with
respect to the static case, up to 
$M_{\rm bar,max}^{({\rm rot})}$. Rotating PNS models with 
$M_{\rm bar,max}^{({\rm stat})}
<M_{\rm bar}< 
 M_{\rm bar,max}^{({\rm rot})}$ are called {\it supramassive}. Such
a supramassive rotating model 
cannot transform into a static model, because in the process of
decreasing angular velocity it will collapse into a black hole. 

%%%%%%%%%%%%%%%%%
\begin{figure}     %3
\epsfig{figure=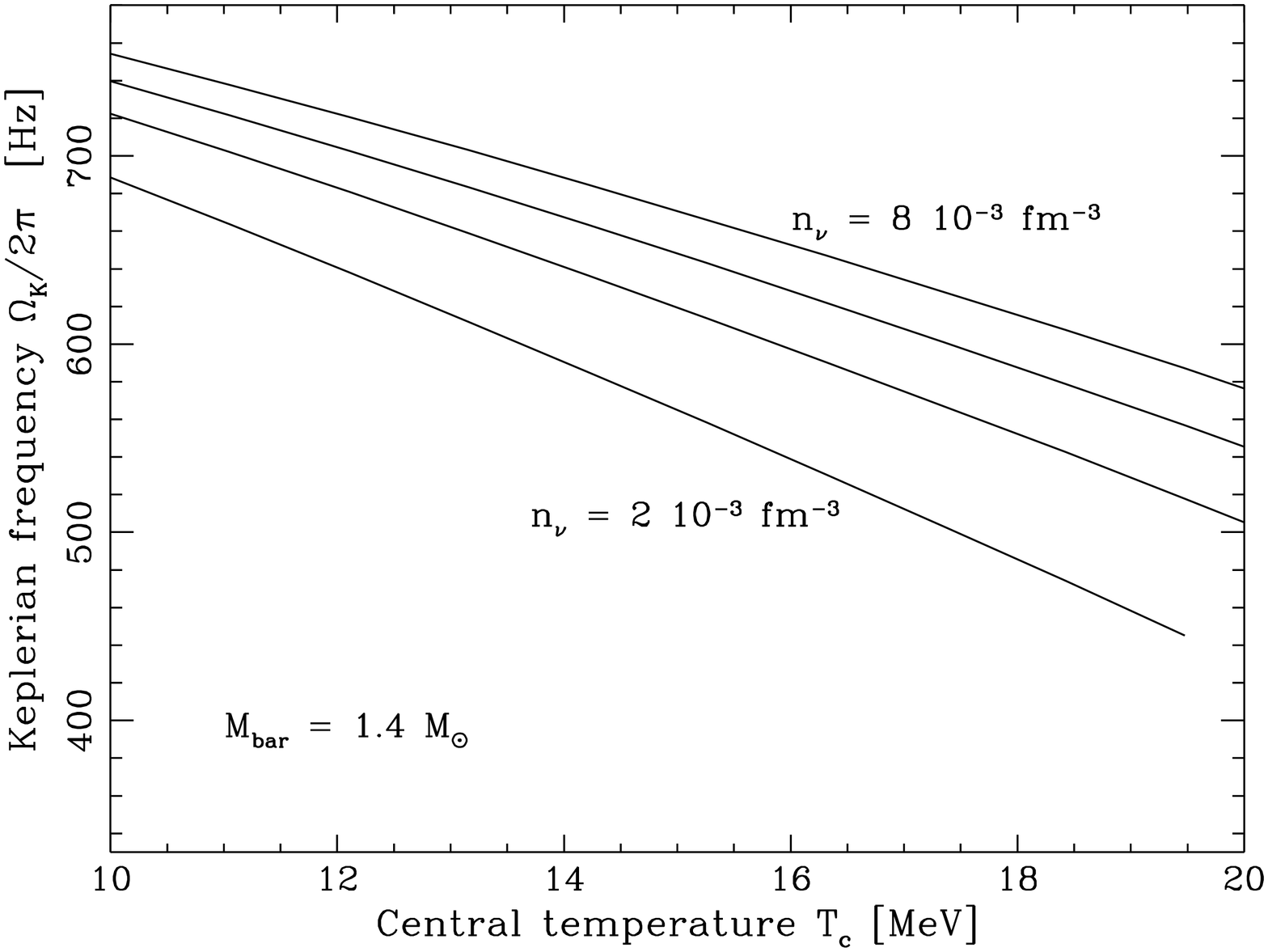,height=8.5cm,angle=-90}
\caption[]{
The value of the Keplerian frequency, $\Omega_K/2\pi$,  versus
central temperature, $T_{\rm c}$, for protoneutron star models
with an isothermal, neutrino opaque, $Y_\nu=0$ (zero trapped
lepton number) hot core. The baryon mass of rotating models is
 fixed at $M_{\rm bar}=1.4~M_\odot$. 
 Different  lines correspond to different
values of the nucleon density at the edge of the hot core,
$n_\nu$. The value of $n_\nu$  increases 
 by $2~10^{-3}~{\rm fm^{-3}}$ when going
upwards from one line to the next one. 
}
\end{figure}
%%%%%%%%%%%%%%%%%%%%%%%%
The maximum angular velocity of the normal and supramassive
rotating models results from different stability conditions. In 
the case of normal rotating configurations the value of $\Omega$
is bound by the mass shedding limit, which corresponds to the
Keplerian velocity at the stellar equator, $\Omega_K(M_{\rm
bar})$. The value of $\Omega_K$ turns out to be rather sensitive to
the location of the ``neutrinosphere'' within the PNS. This
 sensitivity is particularly large in the case of the isothermal
profile of the hot neutrino-opaque core. This is visualized in
Fig. 3, where we show the dependence of the mass shedding limit
$\Omega_K$ for PNS  with baryon mass $1.4~M_\odot$ on the value
of $n_\nu$, for the central temperatures ranging between 10 and
20 MeV. The dependence on the value of $n_\nu$ weakens 
 with increasing $M_{\rm bar}$.
Also,  this effect is much less important in the
case of isentropic PNS.

%%%%%%%%%%%%%%%%%%%%%%%%
\begin{table*}
\centering
\caption[]{
Parameters of the static and rotating  maximum mass configurations of
protoneutron stars
}
%\centering
\begin{tabular}{llllll}
\hline
& & & & &\\
EOS  & $M_{\rm max}^{({\rm stat})}$  & $R_{\rm max}^{({\rm stat})}$  &  
$M_{\rm max}^{({\rm rot})}$ & $R_{\rm max}^{({\rm rot}) \, a}$ 
& $\Omega_{\rm max}/2\pi$ \\
     & $[M_{\odot}]$ & [km] & $[M_{\odot}]$ & [km] & [Hz] \\
& & & & &\\ 
\hline
& & & & & \\
$T=0$, $Y_\nu=0$  & 2.048 & 10.59 & 2.430 & 14.34 & 1625 \\
$T_c=25$, $Y_\nu=0$ & 2.053 & 11.17 & 2.322 & 14.79 & 1521 \\
$s=0.5$, $Y_l=0.4$  & 1.957 & 10.85 & 2.180 & 14.40 & 1522 \\
$s=2$, $Y_l=0.4$  & 1.977 & 11.53 & 2.172 & 15.32 & 1388 \\
& & & & & \\
\hline
& & & & &\\
\multicolumn{5}{l}{$^a$ Equatorial radius of maximaly rotating configuration.}
\end{tabular}
\end{table*}
%%%%%%%%%%%%%%%%%%%%%%
For supramassive rotating PNS, the limit on $\Omega$ has been
set by the condition of stability of rotating models with
respect to the axi-symmetric perturbations, which was combined
with the mass shedding stability condition. 
 Our calculations have been performed for both  isothermal
($T^*=const.$) and isentropic ($s=const.$) hot interiors of PNS.
In both cases, the absolute maximum of Keplerian frequency for
rotating models, which were stable with respect to the
axi-symmetric perturbations, was obtained for a rotating
configuration with a maximum baryonic mass (and gravitational mass), 
$M_{\rm bar,~ max}^{({\rm rot})}$  [$M_{\rm max}^{({\rm rot})}$].
Actually, rotating configuration with $M_{\rm max}^{({\rm rot})}$
and that with $\Omega_{\rm max}$ do not generally coincide (see,
e.g., Cook et al. 1994, Stergioulas \& Friedman 1995). However,
the difference is very small, and it could not  be detected
within the precision of our numerical code. The value 
$M_{\rm max}^{({\rm rot})}$  depends on the value of $T^*$ in the
case of isothermal hot interior, but we preferred to parametrize
it in terms of central temperature,  $T_{\rm c}\equiv T(r=0)$.
%%%%%%%%%%%%%%%%%%%%%%%%
\begin{figure}     %4a
\epsfig{figure=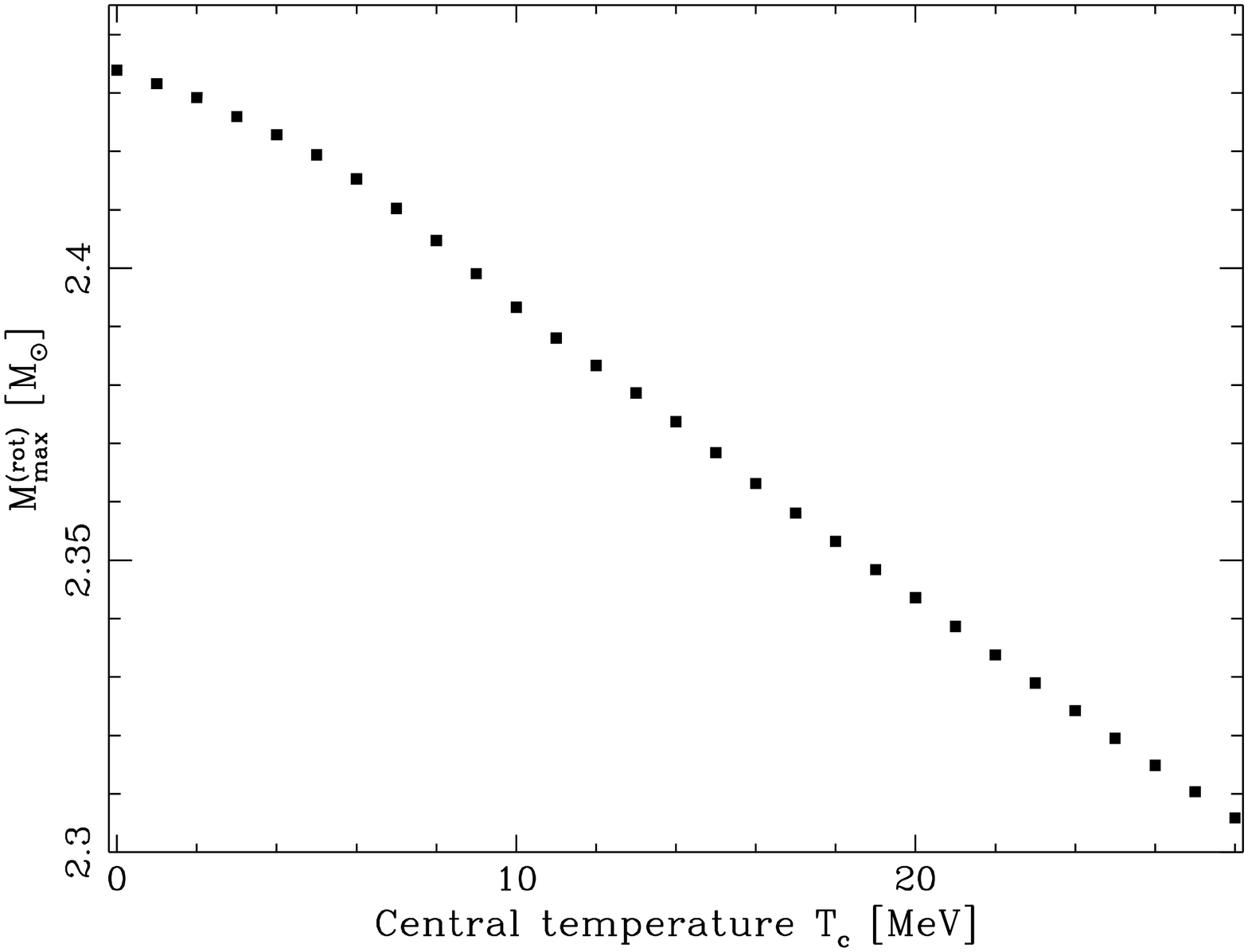,height=8.5cm,angle=-90}
\caption[]{{\bf a.}
Maximum gravitational mass of rotating protoneutron star models with
isothermal cores and $Y_\nu=0$ (zero trapped lepton number),
versus central temperature, $T_{\rm c}$. The nucleon density at
the outer edge of the hot core $n_\nu=5~10^{-3}~{\rm fm^{-3}}$. 
}
\addtocounter{figure}{-1}
%\end{figure}
%%%%%%%%%%%%%%%%%%%%%%%
%\begin{figure}     %4b
\epsfig{figure=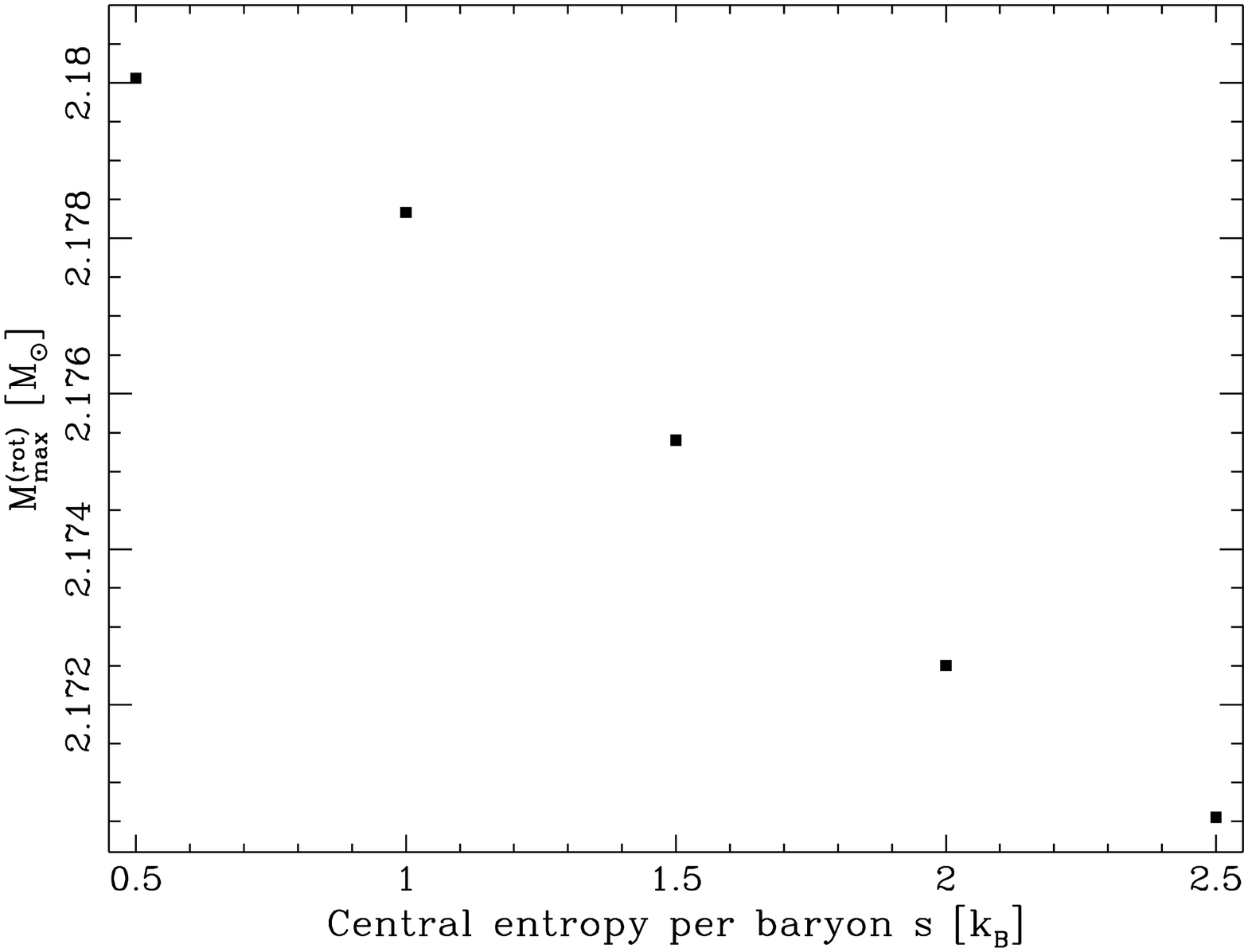,height=8.5cm,angle=-90}
\caption[]{{\bf b.}
Maximum gravitational mass of rotating protoneutron star models with
isentropic hot cores with $Y_l=0.4$, versus central entropy, $s$. 
The nucleon density at
the outer edge of the hot core is 
$n_\nu=5~10^{-3}~{\rm fm^{-3}}$. 
}
\end{figure}
%%%%%%%%%%%%%%%%%%%%%%%%
The dependence of 
$M_{\rm max}^{({\rm rot})}$  on the value of $T_{\rm c}$ is
displayed in Fig. 4 a. The effect of temperature on 
$M_{\rm max}^{({\rm rot})}$  is opposite to that seen for 
$M_{\rm max}^{({\rm stat})}$: thermal effects lower the value 
of $M_{\rm max}^{({\rm rot})}$ as compared to that for cold neutron
stars. This is due to the thermal increase of the  equatorial
radius, which prevails over the thermal stiffening of the
central core of PNS. 
 The dependence of $M_{\rm max}({\rm rot})$  
on the value of $s$ in the case of isentropic hot interior of
PNS is represented in Fig. 4 b, in the case of $Y_\nu \neq 0$ 
(trapped lepton number) and for $Y_l=0.4$. The decrease of 
$M_{\rm max}({\rm rot})$ is there smaller than in the case of
the isothermal PNS models. 

In Fig. 5 a, 5 b we show our results for the maximum rotation
frequency of stable PNS models, reached for 
supramassive configurations.  In the case of
the isothermal PNS, thermal effects tend to decrease the value
of $\Omega_{\rm max}$, but even in the case of $T_{\rm
c}=25~$MeV the relative effect does not exceed ten percent. 
Irregularities at lower $T_{\rm c}$  result from  the 
limited accuracy of determination of the critical configuration
with an isothermal core. 
%%%%%%%%%%%%%%%%%%%%%%%%
\begin{figure}     %5a
\epsfig{figure=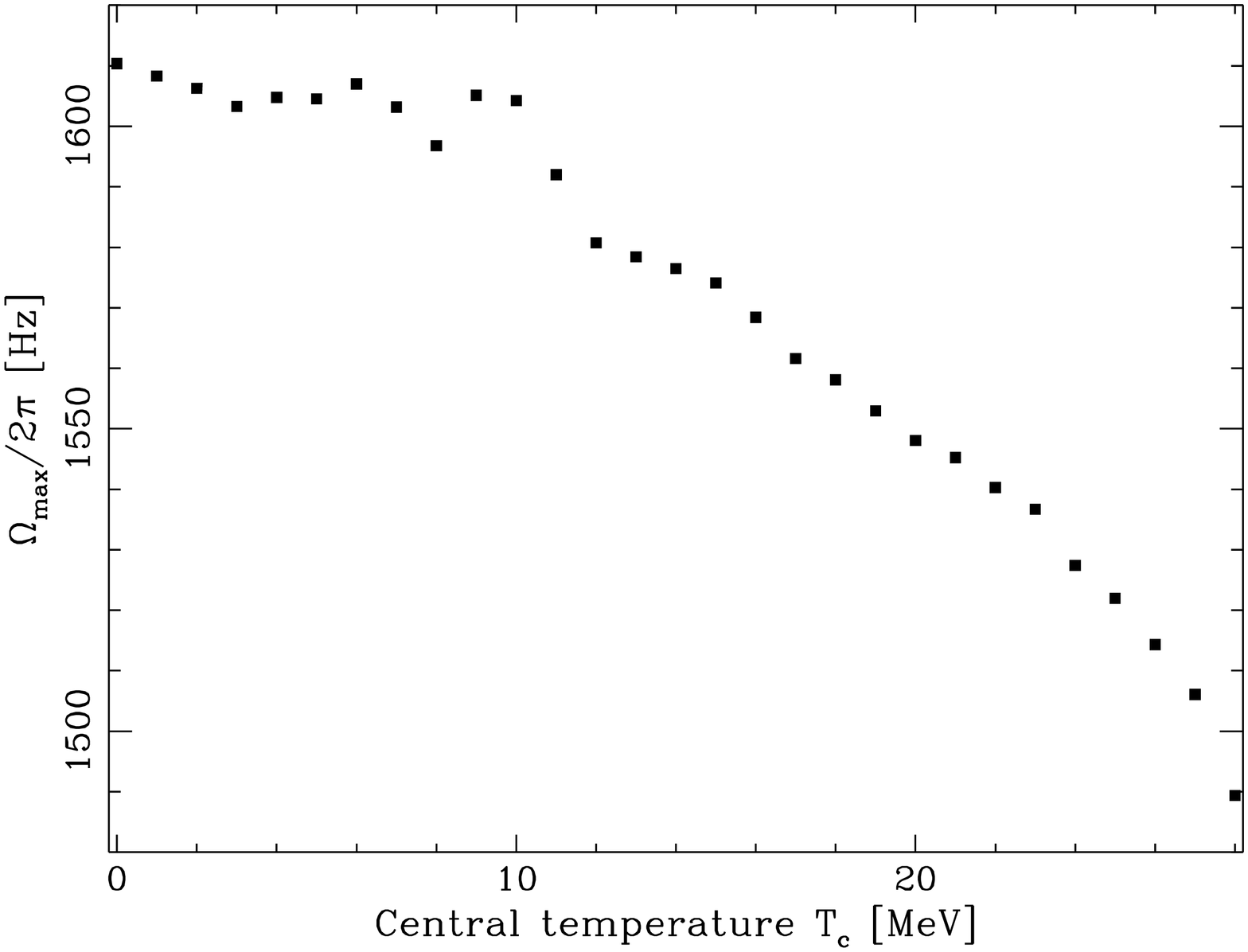,height=8.5cm,angle=-90}
\caption[]{{\bf a.}
Maximum rotation frequency of stable  protoneutron star models with
isothermal hot cores (with zero trapped lepton number), 
versus central temperature, $T_{\rm c}$.
 The nucleon density at the outer edge of the hot core is 
$n_\nu=5~10^{-3}~{\rm fm^{-3}}$. 
}
\addtocounter{figure}{-1}
%\end{figure}
%%%%%%%%%%%%%%%%%%%%%%%%
%\begin{figure}     %5b
\epsfig{figure=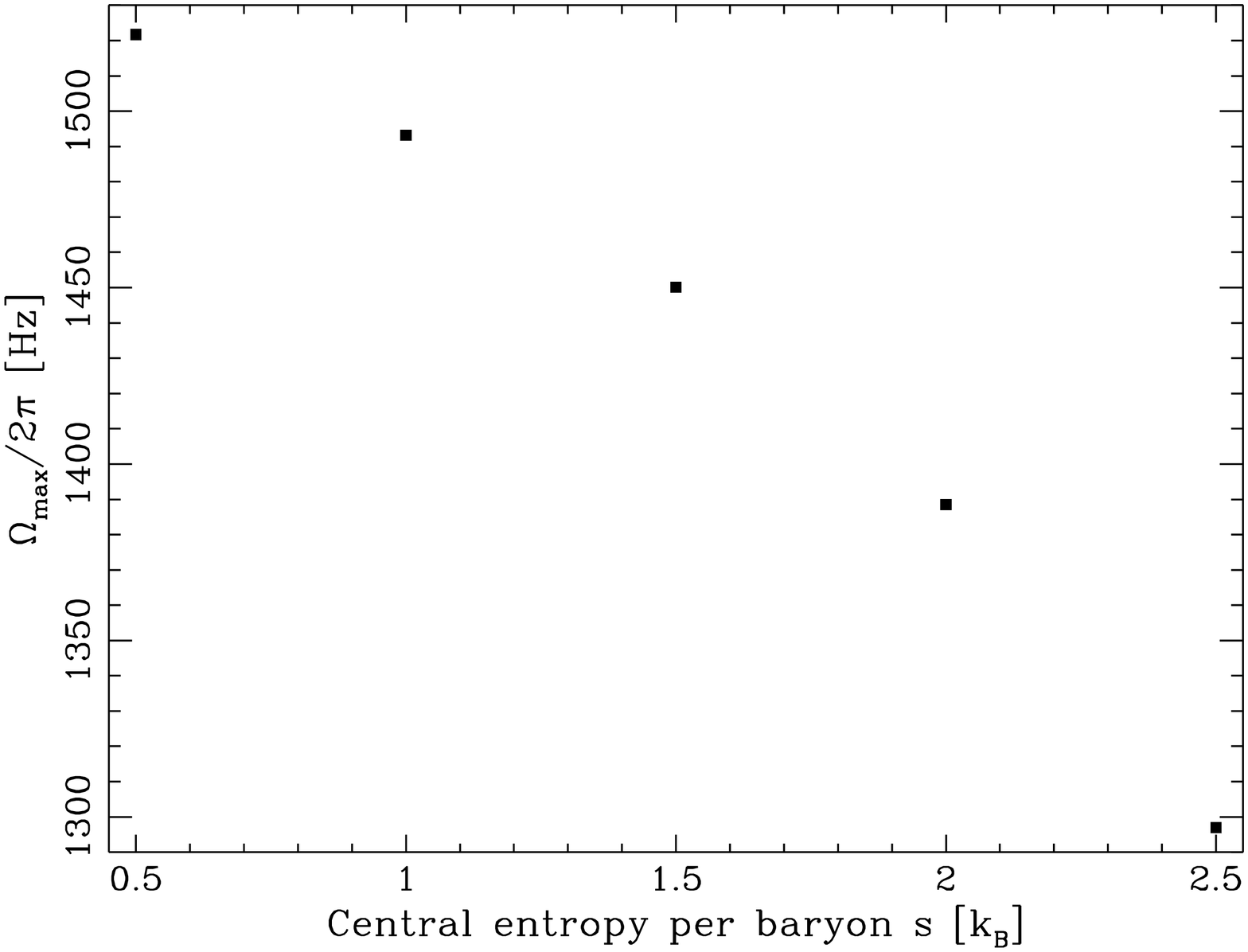,height=8.5cm,angle=-90}
\caption[]{{\bf b.}
Maximum rotation frequency of stable  protoneutron star models with
isentropic hot cores with $Y_l=0.4$, versus central entropy, $s$.  
The nucleon density at
the outer edge of the hot core is 
$n_\nu=5~10^{-3}~{\rm fm^{-3}}$. 
}
\end{figure}
%%%%%%%%%%%%%%%%%%%%%%%%

Maximum rotation frequency in the case of the isentropic hot
neutrino-trapped cores of PNS is shown, for several values of
the central entropy per baryon, in Fig. 5 b.  The value of 
$\Omega_{\rm max}$ is also decreasing with increasing $s$. 

In general, we find that the
decrease of $\Omega_{\rm max}$,  as compared to cold neutron
stars,  is in our case smaller than that found by Hashimoto et al.
(1995); this may be due to the lower value 
of the density at the edge of the hot core, assumed by these
authors. Some differences  may also be due to different
supranuclear EOS, and to a different treatment of the thermal
effects, in particular, to their assumption of $T=const.$. 
%%%%% Rajouts JOG
In order to visualize the importance of the relativistic 
effects on $T(r)$ in the  isothermal interior of a PNS, we show in 
 Fig. 6 the temperatures $T(r)$ and $T^*$,  
 for  $M_{\rm bar}=1.5$ ${\rm M}_{\odot}$, and a 
uniform 
 Keplerian rotation. 
 As one can see, in the central part of
the PNS, $T$ is about $30$ percent greater than $T^{*}$. 
 A similar effect is seen also in the case of an isentropic 
hot interior, which in Fig. 6 corresponds to 
$s=2,~Y_l=0.4$.  
%%%%%%%%%%%%%%%%%%%%%%%%
\begin{figure}     %6
\epsfig{figure=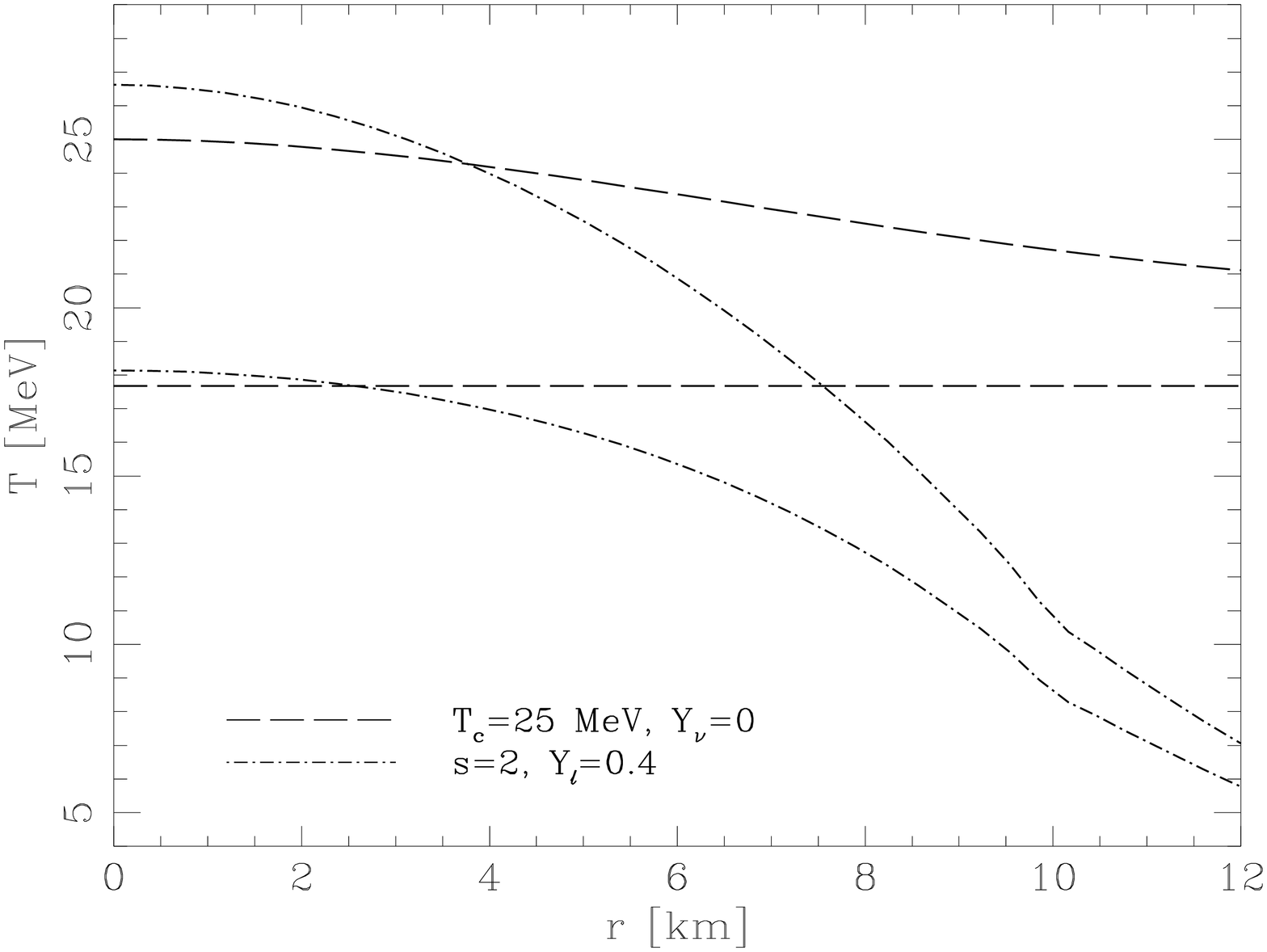,height=8.5cm,angle=-90}
\caption[]{
Difference between $T$ and $T^*$ in the hot interior of
a $M_{\rm bar}=1.5$ ${\rm M}_{\odot}$ PNS in 
Keplerian rotation, for the two models of  
the hot PNS core. Dashed lines correspond to the isothermal 
($T_{\rm c}=15$~MeV), 
and the dash-dot lines to the isentropic model, respectively. 
For each pair of curves, upper one corresponds to $T(r)$, 
and the lower one to $T^*(r)$. Temperatures are plotted 
versus radial coordinate $r$ in the equatorial plane of 
rotating PNS.  
}
\end{figure}
%%%%%%%%%%%%%%%%%%%%%%%%
%%%%% Fin Rajouts
%
%%%%%%%%%%%%%%%%%%%%%%%%%%%%%%%%%%%%%%%%%%%%%%%%%%%%%%%%%
\section{Empirical formula for $\Omega_{\rm max}$}
The calculation of the rotating PNS (and NS) models is
incomparably  more  difficult than that of the static models. 
In the case of cold NS, one finds a surprisingly precise
universal formula, which relates the maximum rotation frequency
to the mass and radius of the maximum mass configuration for the 
static models (Haensel \& Zdunik 1989, Friedman at al. 1989, 
Shapiro et al. 1989, Haensel et al. 1995, Nozawa et al. 1996),
%%%%%%%%%%%%%%%%%
\begin{equation}
\Omega_{\rm max} = 
C \left(
{M_{\rm max}^{({\rm stat})}\over M_\odot}
\right)^{1\over 2}
\left(
{R_{\rm max}^{({\rm stat})}\over 10~{\rm km}}
\right)^{-{3\over 2}}~,
\label{EmpForm}
\end{equation}
%%%%%%%%%%%%%%%
where the most recent value of $C$, based on calculations
performed for a broad set of realistic {\it cold} EOS of dense
matter, is $C_{\rm cold}=7750~s^{-1}$ (Haensel et al. 1995).
The ``empirical formula'', Eq. (10), reproduces the
values of $\Omega_{\rm max}$ for cold NS models with typical
precision better than $5\%$ (although in some specific cases 
the deviations can reach nearly $7\%$, see Nozawa et al. 1996).  

The validity of the empirical formula for $\Omega_{\rm max}$ is of
great practical importance. For cold EOS of dense matter, it
enables one to get immediately a rather precise estimate  of the
value of $\Omega_{\rm max}$, using easily calculated static
neutron star models, and avoiding in this way incomparably more
difficult 2-D 
calculations of rotating NS models and the analysis of their
stability.

%%%%%%%%%%%%%%%%%%%%%%%%
\begin{figure}     %6a
\epsfig{figure=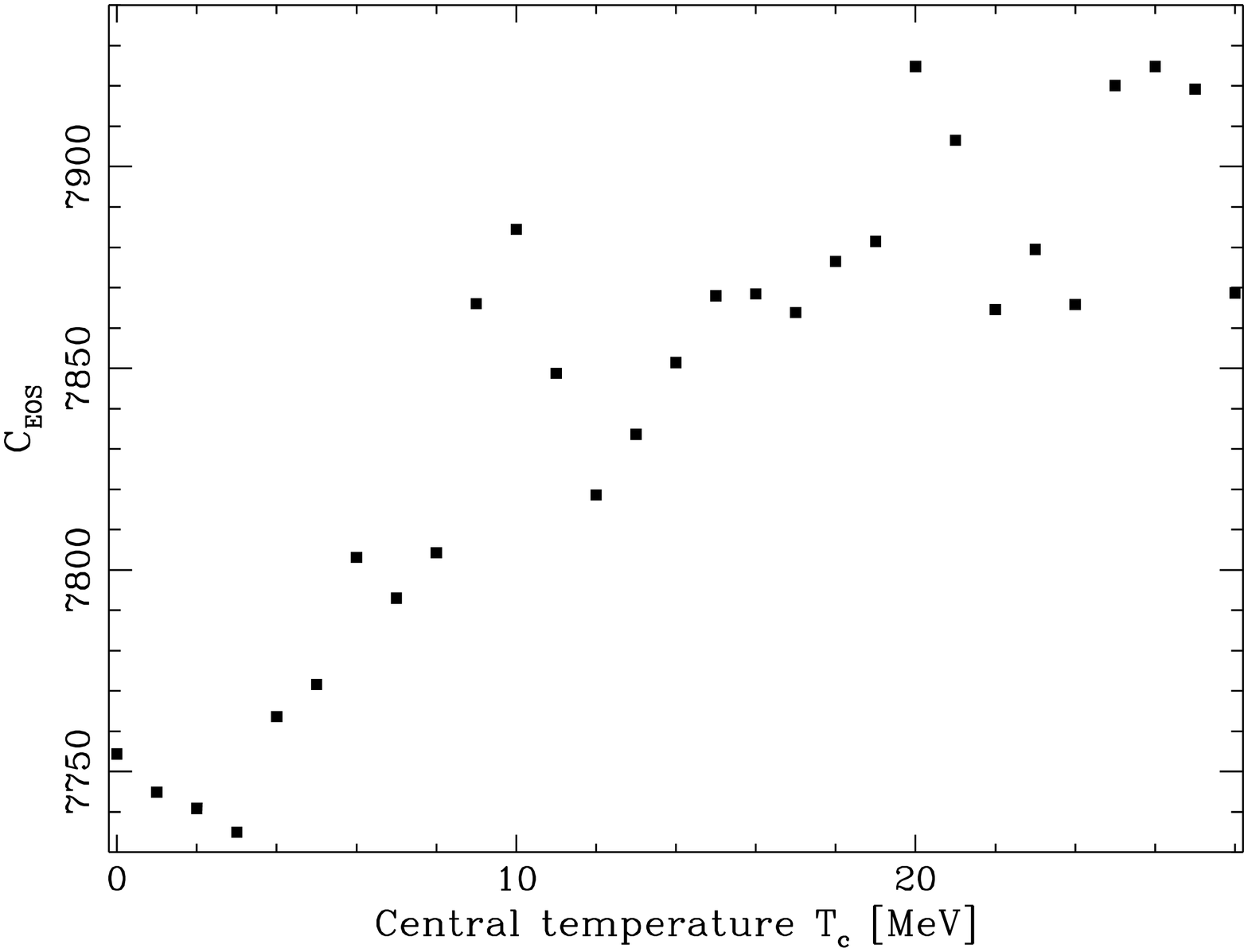,height=8.5cm,angle=-90}
\caption[]{{\bf a.}
The coefficient $C_{\rm EOS}$, given by Eq.(11), versus central
temperature, $T_{\rm c}$, for isothermal models of the zero
trapped lepton number ($Y_\nu=0$) hot cores of protoneutron
stars.  
}
\addtocounter{figure}{-1}
%\end{figure}
%%%%%%%%%%%%%%%%%%%%%%%%
%\begin{figure}     %6b
\epsfig{figure=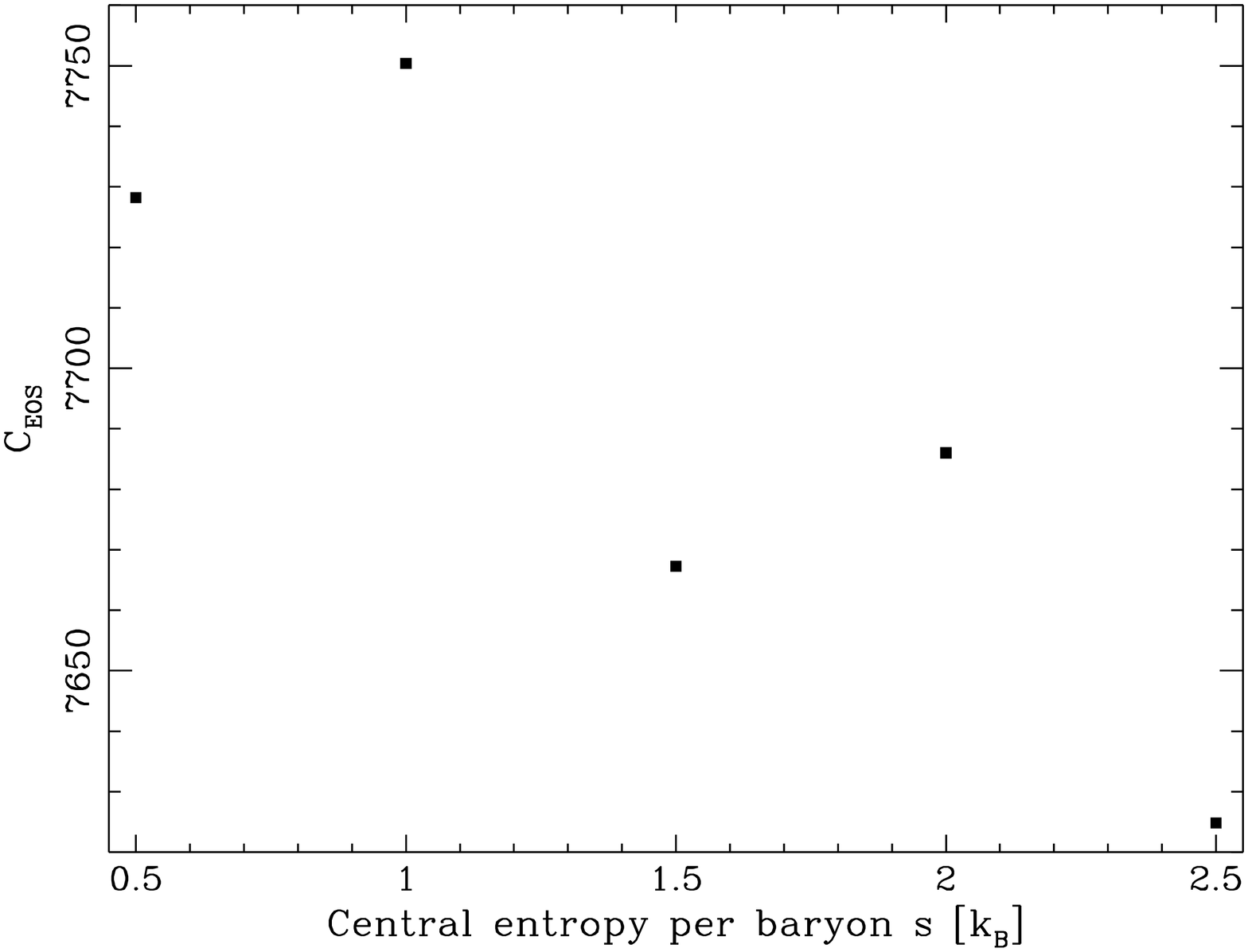,height=8.5cm,angle=-90}
\caption[]{{\bf b.}
The coefficient $C_{\rm EOS}$, given by Eq.(11), versus central
entropy per nucleon,  for isentropic $Y_l=0.4$  models 
of hot cores of protoneutron
stars.  
}
\end{figure}
%%%%%%%%%%%%%%%%%%%%%%%%
It is of interest to check, whether ``empirical formula'' is
valid also in the case of hot PNS. Let us notice, that the
subnuclear EOS of PNS, with possible lepton number trapping, is
very different from that of cold catalyzed matter (see Fig. 1).
In order to check the validity of the empirical formula, we
define the EOS depending parameter  $C_{\rm EOS}$,
%%%%%%
\begin{equation}
C_{\rm EOS}\equiv 
\Omega_{\rm max}
\left(
{ M_{\rm max}^{({\rm stat})}\over M_\odot }
\right)^{-{1\over 2}}
\left(
{ R_{\rm max}^{({\rm stat})}\over 10~{\rm km} }
\right)^{3\over 2}~,
\label{CEOS}
\end{equation}
%%%%%%%%%%
where the right-hand-side is calculated using exact results for a
specific EOS. The values of $C_{\rm EOS}$ for our  models
of PNS are displayed in Fig. 7 a, 7 b. Let us notice, that for
isentropic hot cores with trapped lepton number 
the value of $C_{\rm EOS}$ shows a decreasing trend with
increasing $s$. In the 
case of isothermal hot cores with $Y_\nu=0$ we see the opposite
trend: the value of $C_{\rm EOS}$ tends to increase with central
temperature. The value of $C_{\rm hot}=7800~{\rm s^{-1}}$
will lead to a very good empirical formula for PNS, with
precison comparable to that used for cold NS. Within the
precision of the empirical formulae, the values of 
$C_{\rm cold}$ and $C_{\rm hot}$ can be considered as being equal. 
%
%%%%%%%%%%%%%%%%%%%%%%%%%%%%%%%%%%%%%%%%%%%%%%%%%%%%%%%%%
\section{Numerical results. From hot protoneutron stars to cold
neutron stars} 

\begin{table}
\caption[]{
Maximum rotation frequencies and maximum baryon masses of PNS
with various EOS, and corresponding frequencies of cold NS
configurations, obtained via cooling at constant $J$ and
$M_{\rm bar}$ ($J$-losses due to neutrino
emission are neglected). 
}
\centering
\begin{tabular}{llll}
\hline
 & & &\\
EOS  & $M_{\rm bar,max}^{({\rm rot})}$ & 
$\Omega_{\rm max}^{({\rm hot})}/2\pi$ &  
$\Omega_{\rm max}^{' ({\rm cold})}/2\pi$\\ 
     & $[M_{\odot}]$ & [Hz] & [Hz] \\
 & & &\\     
\hline
 & & &\\
$T_c = 9$, $Y_\nu = 0$  & 2.806 & 1603 & 1582 \\
$T_c = 25$, $Y_\nu = 0$ & 2.695 & 1521 & 1403 \\
$s = 2$, $Y_L = 0.4$ & 2.172 & 1388 & 1112 \\
 & & &\\
\hline
\end{tabular}
\end{table}
\vspace{0.5cm}
%%%%%%%%%%%%%%%%%%%%%%%%
Rapidly rotating hot, neutrino-opaque PNS evolves eventually into
a cold, rotating  NS, which under favourable circumstances can be
observed as a solitary pulsar. Let us assume, that such a
transformation took place at constant baryon mass:
 this assumption is valid if mass accreted after the formation
of a PNS is negligibly small. 
(Notice, that we can {\it define} the moment of formation of a
PNS as that at which the accretion ends). Second assumption refers to the
angular momentum of the star. Let us neglect for the time being
the angular momentum loss due to emission of neutrinos (mainly
during deleptonization), as well as that due to gravitational
radiation (if rotating star deviated from the axial symmetry).
Inclusion of both effects could only {\it decrease} the final angular
momentum of the NS. Under these two assumptions, the transformation
of a PNS into a solitary radio pulsar takes place at constant
baryon mass, $M_{\rm bar}$, and angular momentum, $J$. 
The maximum angular frequency of a solitary radio pulsar of a given baryon
mass would be then determined by the parameters of a maximally
rotating PNS  of the same baryon mass. 

In the present paper we restrict ourselves, similarly as 
Hashimoto et al. (1995), to the case of rigid
rotation of PNS. We will first consider ``typical'' range of baryon
masses expected for the NS born in a gravitational collapse of a
massive stellar core, $1.4~M_\odot < M_{\rm bar} < 2.0~M_\odot$
(for our EOS of dense matter this would correspond to 
$1.3~M_\odot \widetilde{<} M \widetilde{<} 1.8~M_\odot$). 
%\footnote{About possibility of  
%$M_{\rm max}\simeq 1.5~M_\odot$ 
% advanced by Bethe \& Brown 1994}. 
All stars considered are normal, and the
maximum initial frequency of PNS will be then 
$\Omega_K[{\rm hot}, Y_\nu=0]$, and
$\Omega_K[{\rm hot}, Y_l=0.4]$, for the limiting
case of the zero trapped lepton number, and maximum trapped
lepton number neutrino-opaque cores, respectively. The values of
these limiting angular frequencies, for the isothermal and
isentropic PNS cores, are plotted versus $M_{\rm bar}$ in Fig.
8 (thin dashed and dash-dotted lines, respectively).  

%
%%%%%%%%%%%%%%%%%%%%%%%%
\begin{figure}     %7
\epsfig{figure=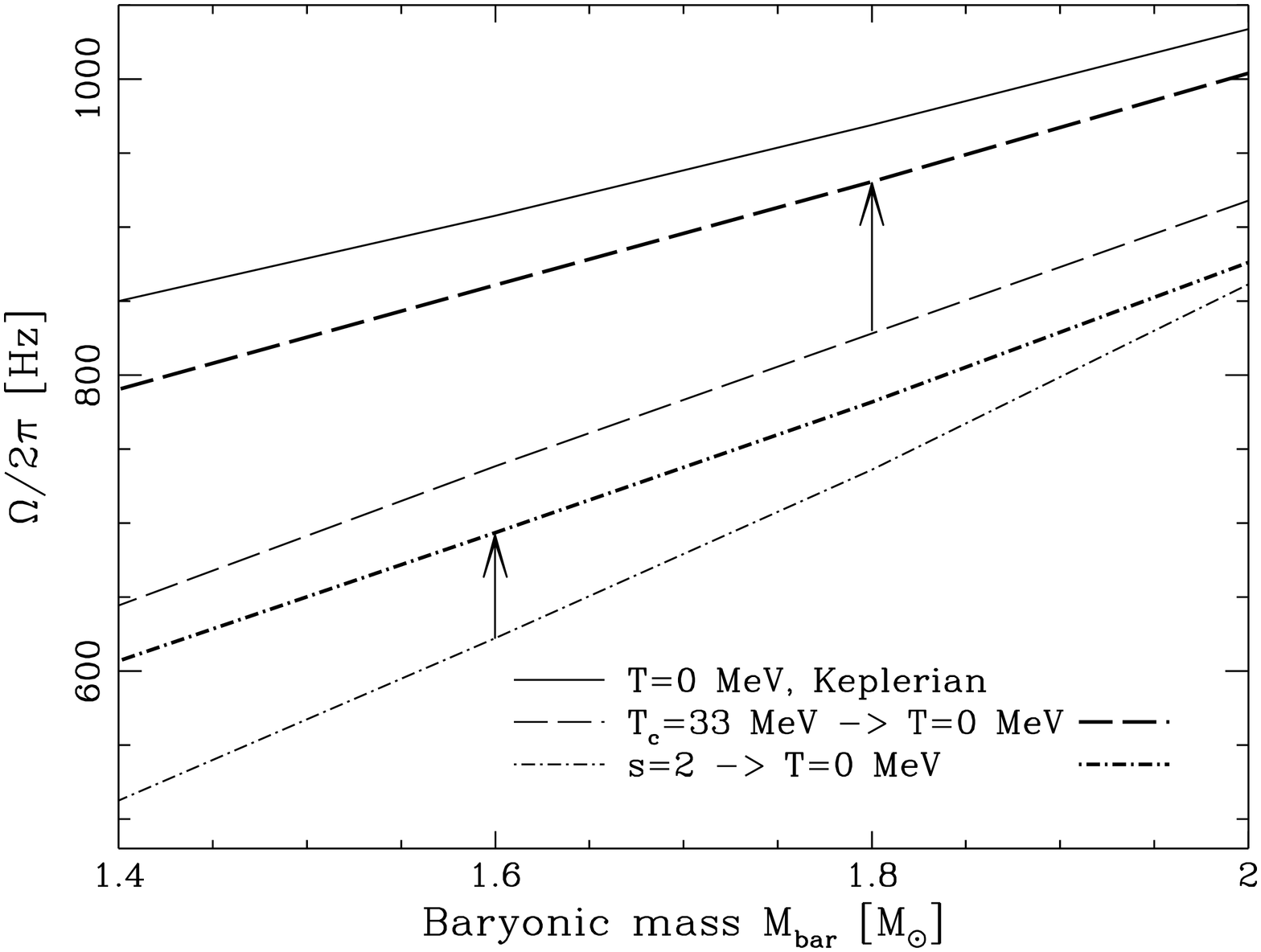,height=8.5cm,angle=-90}
\caption[]{
Maximum rotation frequencies versus baryon mass of rotating
configurations. 
Solid line: maximum rotation frequencies of cold neutron star
models, $\Omega_K({\rm cold})$.  Thin dashed line: maximum
rotation frequency for  isothermal models with $T_{\rm
c}=33~$MeV with no trapped lepton number ($Y_\nu=0$), denoted in
the text as $\Omega_K[{\rm hot},Y_\nu=0]$. Heavy dashed 
line: angular frequency of cold configurations, resulting from 
those corresponding to thin dashed line by cooling at
constant $J$ and $M_{\rm bar}$. 
Thin dash-dotted line: maximum
rotation frequency for  isentropic  models, denoted in
the text as $\Omega_K[s=2,Y_l=0.4]$. Heavy dash-dotted 
line: angular frequency of cold configurations, resulting from 
those corresponding to thin dash-dotted line by cooling at
constant $J$ and $M_{\rm bar}$ ($J$-losses due to neutrino
emission are neglected). 
}
\end{figure}
%%%%%%%%%%%%%%%%%%%%%%%%

Our results displayed in Fig. 8 lead to some  conclusions,
relevant to those neutron stars which remained solitary after
their birth in the gravitational collapse of a massive stellar
core. If the starting configuration (at the end of significant
accretion, after the revival of a stagnated shock)
 was that with trapped lepton
number and isentropic, then the maximum frequency of such a
neutron star, in the gravitational mass range of 1.3 -
1.8~$M_\odot$, cannot exceed 600 - 900~Hz, the lower  frequency
limit  corresponding to the lower mass limit.
The corresponding range of minimum periods would be  from
1.6~ms to  1.1~ms, respectively. 

If the ``initial isolated
configuration'' (i.e., that with no subsequent accretion) was a
little older, and therefore deleptonized, and if we assume 
 that it
was isothermal, then the resulting
constraints would be less stringent: for the range of masses 1.3
- 1.8~$M_\odot$ we get minimum rotation periods of 1.25 - 1.1~ms,
respectively.  

In both cases cooling resulted in speeding up of the rotation of
the star: such situation is characteristic of ``normal rotating
models''. 

Similar evolutionary considerations can be applied to the ``maximally
rotating configurations'': isentropic one  with $s=2,~Y_l=0.4$
and  isothermal ones with $T_{\rm
c}=9~$MeV,~25~MeV, respectively. Our results are shown in
Table 2. The initial ``hot'' configurations are supramassive,
and evolve into more compact cool ones  (at constant $M_{\rm
bar}$ and $J$)  {\it decreasing} their angular frequency (slowing down). 
This purely
relativistic effect was recently pointed out by Hashimoto et al.
(1995).  The ``relativistic slowing down'' during cooling of the
maximally rotating isentropic configuration with $s=2,~Y_l=0.4$
corresponds to the increase of the period from 
$P_{\rm min}[s=2,Y_l=0.4]=0.72$~ms  up to $P'_{\rm min}=0.90$~ms. 
If such
a scenario of formation of solitary pulsars is valid, an
 absolute limit on their period would be $P'_{\rm min}$, and not 
$P_{\rm min}(T=0)=0.61$~ms, obtained for the $T=0$ EOS. The
slowing down factor of $\simeq {2\over 3}$ coincides with that
obtained by Hashimoto et al. (1995) for a different EOS and a
different scenario of formation of solitary cold pulsars.
Actually, they assumed that the initial
configuration is that with $T=const., Y_\nu=0$. Their initial
model  would to some extent correspond to our isothermal models
(remember however the factor ${N\over \Gamma}$ in our
temperature profiles, which is absent in their calculations). 
Using Tables 1, 2 we obtain 
$P'_{\rm min}[T_c=25,Y_\nu=0]/P_{\rm min}[T=0]=0.92$, which is 
quite close to 1. 
%
%%%%%%%%%%%%%%%%%%%%%%%%%%%%%%%%%%%%%%%%%%%%%%%%%%%%%%%%%
\section{Discussion and conclusions}
In the present paper we studied rapid rotation of protoneutron
stars, using a specific model of hot, dense matter.  
Our models of rapidly rotating protoneutron stars were based on
several simplifications, which were necessary in order to make
the problem tractable. 
In order to avoid difficulties and/or ambiguities of the largely
unknown ``real situation'', we
restricted ourselves to studying idealized, limiting cases.
In many places we introduced approximations, which were crucial
for making numerical calculations feasible.

We assumed rigid rotation within protoneutron star. This
simplified greatly our considerations, and reduced dramatically the
number of stellar models. Actually, our formalism allows calculation of
the quasi-stationary  differentially rotating configurations,
after introducing an additional   
$F\partial_i\Omega$ term appearing on the
right hand side of the equation of stationary  motion, Eq. (4)
 (see Section 5.1 of BGSM). 
We plan to perform studies of differentially rotating
protoneutron stars in the near future, as the next step in 
our investigations of dynamics of protoneutron stars. 
 While existing numerical simulations of gravitational collapse of 
 rotating cores of massive stars yield a differentially rotating 
 protoneutron star, the calculations stop too early after 
 bounce. In contrast to our models of protoneutron stars,  the 
 object which comes out from the simulations of Moenchmeyer and 
 Mueller has an extended, very hot envelope, produced by the 
 shock wave immediately after bounce (Janka \& Moenchmeyer
 1989a,b, Moenchmeyer \& Mueller 1989). Also, the lack of 
 knowledge of the initial angular velocity distribution within the 
 collapsing core results in the uncertainty in the 
 rotational state of produced protoneutron star. Clearly, 
 the study of quasistationary differential rotation of protoneutron 
 stars should take into account all these uncertainties.

Deleptonization of protoneutron star is connected with energy
 and angular momentum losses. However, angular momentum taken
away by neutrinos is expected  to constitute at most a few
percent of the  total stellar angular momentum (Kazanas 1977).
Inclusion of this 
effect would slightly  decrease some of our ``evolutionary
limits'' of Section 7. 

Our treatment of the thermal state of the  protoneutron star
interior should be considered as very crude. The temperature
profile might be affected by convection. Also, our method of
locating the ``neutrinosphere'' was very approximate. Clearly, the
treatment of thermal effects can be refined, but we do not think
this will change our main results. 

Our calculations were performed for only one model of the
nucleon component of dense hot matter. The model was realistic,
and enabled us to treat in a unified way the whole interior
(core as well as  the envelope) of
the protoneutron star. However, in view of the uncertainties in
the EOS of dense matter at supranuclear densities, one should of
course study the whole range of theoretical possibilities, for a
broad set - from soft to stiff - of supranuclear, high
temperature EOS.  An example of such an investigation, 
in the case of the
{\it static} protoneutron stars, is the study of Bombaci et al.
(1996). In view of the possible importance of the protoneutron
star - neutron star connection for the properties of solitary
pulsars, similar studies should also be done for rotating
proto-neutron stars. 

The calculations of the present paper were done under the
assumption, that the frequency of uniform rotation of
protoneutron stars is limited only by the mass shedding and the secular
instability with respect to the axi-symmetric perturbations. 
In view of this, our results can be considered only as upper
bounds to the maximum rotation frequency. Our main conclusion is
that the minimum rotation period of solitary neutron stars,
born as rapidly rotating protoneutron stars, is 
significantly  larger,  than the corresponding limit for cold
neutron stars.   
 Inclusion of additional secular instabilities in rapidly, uniformly 
rotating protoneutron stars can only strengthen this conclusion.

\begin{acknowledgements} We are very grateful to E. Gourgoulhon
for his help at the initial stage of this project. 
We are also very grateful to W. Dziembowski for introducing us 
into the difficult subject of meridional circulation.  
This research was partially supported by the JUMELAGE program 
``Astronomie France-Pologne'' of CNRS/PAN, by the KBN grant No. P304 014 07,
and by the MESR grant no. 94-3-1544. 
The numerical computations have been performed on the Silicon 
Graphics workstations, purchased thanks to the support of 
the SPM department of the CNRS and the Institut National des 
Sciences de l'Univers.  
\end{acknowledgements}

\end{document}